\documentclass[a4paper,aps,prb,twocolumn] {revtex4}
\usepackage{amsmath}
\usepackage{graphicx}
\usepackage{subfigure}

\def \be{\begin{equation}\begin{aligned}}
\def \ee{\end{aligned}\end{equation}}
\def \q{\vec q}
\def \o{\omega}
\def \s{\sigma}
\def \mD{\mathcal D}

\def \e{\epsilon}
\def \la{\langle}
\def \ra{\rangle}
\def \d{\mathrm{d}}

\DeclareMathOperator{\im}{Im}

\begin{document}
\title{Clausius-Clapeyron Relations For First Order Phase Transitions \\ In Bilayer
Quantum Hall Systems}
\author{Yue Zou${}^1$, Gil Refael${}^1$, Ady Stern${}^2$ and J. P. Eisenstein${}^1$}
\affiliation{${}^1$Department of Physics, California Institute of
Technology, Pasadena, California 91125, USA \\ ${}^2$Department of Condensed Matter Physics, Weizmann Institute of Science, Rehovot 76100, Israel}
\date{\today}

\begin{abstract}
A bilayer system of two-dimensional electron gases in a perpendicular magnetic field exhibits rich phenomena. At total filling factor $\nu_{tot} = 1$, as one increases the layer separation, the bilayer system goes from an interlayer coherent exciton condensed state to an incoherent phase of, most likely, two decoupled composite-fermion Fermi liquids. Many questions still remain as to the nature of the transition between these two phases.  Recent experiments have demonstrated that spin plays an important role in this transition. Assuming that there is a direct first order transition between the spin-polarized interlayer-coherent quantum Hall state and spin-partially-polarized composite Fermi liquid state, we calculate the phase boundary $(d/l)_c$ as a function of parallel magnetic field, NMR/heat pulse, temperature, and density imbalance, and compare with experimental results. Remarkably good agreement is found between theory and various experiments.
\end{abstract}

\maketitle

\section{Introduction}

A bilayer two dimensional electron gas in a strong perpendicular magnetic field at total filling factor $\nu_{tot}=1$ exhibits rich phenomena. An important tuning parameter in this system is the ratio $d/l$, where $d$ is the effective interlayer distance, and $l$ is the magnetic length. At small $d/l$, even with negligible tunneling, a remarkable bilayer quantum Hall state with interlayer phase coherence emerges due to interlayer Coulomb interaction. This bilayer quantum Hall state can be described as a pseudospin ferromagnet where the layer index acts as the pseudospin, or an exciton condensate formed by interlayer particle-hole pairing\cite{WenZee1993,Moon1995,Eisenstein2004}. Many remarkable experimental signatures of this phase predicted by theories have been observed in experiments, including enormous enhancement of zero bias interlayer tunneling\cite{Spielman2000}, linearly dispersing Goldstone mode\cite{Spielman2001}, quantized Hall drag\cite{Kellogg2002}, and vanishing resistance in counter flow\cite{Kellogg2004}. However, there are still important discrepancies between theory and experiment. For example, the height of the interlayer tunneling conductance is observed to be finite\cite{Spielman2000}, while theories predict it to be infinite. Also, transport in counterflow experiments should be completely dissipationless under a critical temperature for phase coherence, but in experiments dissipationless counterflow is only seen in the zero-temperature limit\cite{Kellogg2004}. The effect of quenched disorder is believed to be crucial to reconcile these discrepancies\cite{Stern2001,MacDonald2001,Fertig2005}, although a quantitative understanding is still lacking.

The nature of the phase transition when $d/l$ is increased and the quantum Hall phase is destroyed is even more puzzling. In the limit $d/l\rightarrow\infty$, each layer is at half-filling, and they should behave as weakly-coupled composite Fermi liquids. Much progress have been made in understanding this phase using the Chern-Simons approach\cite{HalperinLeeRead,MRPA,SternHalperin1995,MMRPA,Simon1998} and the dipolar quasiparticle approach\cite{Shankar1997,Read1998,Lee1998,Stern1999,Shankar2001,Shankar2003}. Although we understand well both the coherent phase at $d/l\rightarrow0$ and the composite Fermi liquid state at $d/l\rightarrow\infty$, the transition between them has been shrouded in mystery. There have been many experimental\cite{Kellogg2003,Tutuc2003,Spielman2004,Clarke2005,Spielman2005,Kumada2005,Luin2005,Tracy2006,Fujisawa2008,Champagne2008,Champagne2008b,Karmakar2009,Finck2010,Muraki2010} and theoretical\cite{Cote1992,ZhengFertig1995,Ho1995,Bonesteel1996,Kim2001,Demler2001,Schliemann2001,SternHalperin2001,Simon2003,Shibata2006,Ye2007,Moller2008a,Moller2008} studies regarding the nature of this transition. While some of these theoretical works point to a direct transition between the two limiting phases, either continuous\cite{Shibata2006} or of first order\cite{Schliemann2001,SternHalperin2001}, some other works predict the existence of various types of exotic intermediate phases, including translational symmetry broken phase\cite{Cote1992,ZhengFertig1995,Ho1995,Ye2007}, composite fermion paired state\cite{Bonesteel1996,Kim2001,Moller2008a}, phase of coexisting composite fermions and composite bosons\cite{Simon2003,Moller2008,Papic2009}, and quantum disordered phases\cite{Demler2001}, etc.

These theoretical works typically assume that the physical spin is
fully polarized and hence irrelevant across the transition. However,
recent experiments have shown that spin plays an important role in the
transition. Ref. \onlinecite{Spielman2005} has found that by applying
a NMR pulse or heat pulse to depolarize the nuclei and hence
increasing the effective magnetic field coupled to the spin, the
coherent phase is strengthened, and the phase boundary shifts to
higher value of $d/l$. Similar behavior has also been observed by
applying a parallel magnetic
field\cite{Fujisawa2008}. These experimental results indicate that at least one of the
phases involved in the transition is not fully polarized, and that the
polarization changes significantly accross the transition. The most
likely possibility is that the incoherent composite Fermi liquid phase at large
$d/l$ is only partially polarized, as shown by other experiments on single layer at
$\nu=1/2$\cite{Tracy2007,Li2009}. If the transition between the
coherent phase and the {\it less polarized} incoherent phase is a
thermodynamic phase transition, it must be of first order: The
magnetization is discontinuous across the transition, and, as the
experiments of Ref. \onlinecite{Spielman2005} found, the
transition can be tuned using a Zeeman field which is conjugate to the
magnetization. These two facts together imply the first order nature
of the transition. An alternative to the thermodynamic transition
scenario is a singularity-free quantum crossover as was suggested
recently in Refs. \onlinecite{Moller2008a,Moller2008}.

In this work, we assume that the transition tuned by $d/l$ is a
thermodynamic first-order transition between spin-polarized coherent $\nu_{tot}=1$
quantum Hall state and partially-polarized composite Fermi liquid
state, and derive the Clausius-Clapeyron relations for this
system. The Clausius-Clapeyron relations will allow us to obtain the
phase boundary shapes for the transition; a comparison of these
boundaries with experiments presents a stringent consistency test of
the first order transition scenario. The first-order scenario was invoked
by Ref. \onlinecite{SternHalperin2001} to explain the strongly
enhanced longitudinal Coulomb drag for intermediate $d/l$, and it also
has some support from exact-diagonalization
study\cite{Schliemann2001}. Note that we will only consider the case of negligible interlayer tunneling.

The Clausius-Clapeyron relations are the results of matching the free
energies of the two phases along the phase boundary.  To be more specific, we denote the free energy density of the coherent and the incoherent phases to be $E_c$ and $E_i$, and define
\be
f(\delta,B_{tot},\Delta n, T) = E_c(\delta,B_{tot},\Delta n, T) - E_i(\delta,B_{tot},\Delta n, T),\nonumber
\ee
where $\delta\equiv d/l$, $B_{tot}$ is the total magnetic field coupled to electrons' physical spin, $\Delta n = (n_1 - n_2) / 2$ is the density imbalance, $T$ is the temperature. At any point along the phase boundary, we must have
\be
f(\delta_c,B_{tot},\Delta n, T) =0.
\ee
This equation can be viewed as defining the $\delta_c$ at which the
transition occurs. When one changes the total field by $\d B_{tot}$, the critical $\delta_c(B_{tot},\Delta n, T)$ also changes by $\d\delta_c$ when the filling factor is kept fixed at $\nu_{tot}=1$. Their relation is determined by
\be
0 = \frac{\partial f}{\partial \delta}\d\delta_c + \frac{\partial f}{\partial B_{tot}}\d B_{tot},
\ee
therefore the slope of the phase boundary is determined by the following ODE:
\be \label{spin}
\frac{\d\delta_c}{\d B_{tot}}=-\frac{\frac{\partial f}{\partial B_{tot}}}{\frac{\partial f}{\partial\delta}}=\frac{\frac{\partial E_i}{\partial B_{tot}}-\frac{\partial E_c}{\partial B_{tot}}}{\frac{\partial f}{\partial\delta}}.
\ee
A crucial assumption of our work is that
\be\label{eta}
\frac{\partial f}{\partial\delta}=\eta\frac{e^2}{\epsilon l^3},
\ee
where $e^2/(\epsilon l^3)$ not only gives the correct units, but is
the only energy scale that exists in this problem if we neglect
the Landau Level mixing. $\eta$ is a universal positive dimensionless constant. It is positive because $f$ should be an increasing function of $\delta=d/l$, since the incoherent phase should be more and more energetically favorable with increasing $d/l$. In general, $\eta$ could be a function of $\delta=d/l$, i.e., $\eta(\delta)\approx\eta(\delta_0)+\mathcal O[(\delta-\delta_0)/\delta_0]$, but since in experiments $\delta$ does not change much (ranging from $1.7$ to $2$), $(\delta-\delta_0)/\delta_0\ll1$, we will assume $\eta$ to be a constant for simplicity.

Similar analysis also applies to finite temperature transitions:
\be \label{finiteT}
\frac{\d\delta_c}{\d T}=\frac{\frac{\partial E_i}{\partial T}-\frac{\partial E_c}{\partial T}}{\eta\frac{e^2}{\epsilon l^3}}.
\ee
For density imbalance experiments, we will focus on the phase boundary near $\Delta n =0$. First, note that by symmetry
\be
\frac{\partial f}{\partial \Delta n} = 0.
\ee
Thus, we need to expand $f$ to second order in $\Delta n$:
\be
0 = \frac{\partial f}{\partial \delta}\d\delta_c +\frac12 \frac{\partial^2 f}{\partial \Delta n^2}(\Delta n)^2,
\ee
and therefore
\be \label{imbalance}
\frac{\d\delta_c}{\d (\Delta n^2)}=\frac{\frac12\frac{\partial^2 E_i}{\partial \Delta n^2}-\frac12\frac{\partial^2 E_c}{\partial \Delta n^2}}{\eta\frac{e^2}{\epsilon l^3}}.
\ee

The above equations constitute the Clausius-Clapeyron relations for the bilayer quantum Hall systems. In the following sections, we will investigate whether the phase boundary shapes implied by Clausius-Clapeyron relations are consistent with experiments, and whether a single universal parameter $\eta$ can explain all available experimental results. To obtain the detailed forms of free energy of both phases, we will primarily work with the pseudospin ferromagnet description for the coherent quantum Hall phase and the Chern-Simons approach for the incoherent composite Fermi liquid phase. Spin transitions, finite temperature transitions, and density imbalance experiments are studied in Sec. \ref{sec:magnetic}, \ref{sec:finiteT}, and \ref{sec:imbalance}, respectively. Finally, we summarize and discuss our results in Sec. \ref{sec:summary}. Some theoretical details are relegated to Appendices.

\section{Spin transition experiments}\label{sec:magnetic}
Ref. \onlinecite{Spielman2005} and Ref. \onlinecite{Fujisawa2008} have
studied the effect of NMR/heat pulse and parallel magnetic field on
the transition tuned by $d/l$, respectively. In the experiment of
Ref. \onlinecite{Fujisawa2008}, since the interlayer tunneling is
negligible, the main effect of the parallel field is on the spins of
electrons. Similarly, in the experiment of
Ref. \onlinecite{Spielman2005}, NMR/heat pulse acts to depolarize the
nuclei and therefore also changes the Zeeman field on the electrons
through the hyperfine coupling. Thus, these two experiments can be analyzed in a similar fashion. Since we assume the coherent phase is spin polarized, the spin part of the coherent phase free energy is simply the Zeeman energy:
\be
E_c&=-\frac12N_T|g|\mu_BB_{tot}=-\frac{e|g|\mu_BB_{\perp}B_{tot}}{4\pi\hbar},
\ee
where $N_T$ is the total electron density of the two layers, $B_{\perp}$ is the perpendicular magnetic field, $B_{tot}$ is the total magnetic field coupled to electron spin, $g=-0.44$ is the $g$-factor of the GaAs two dimensional electron gas, and $\mu_B$ is the Bohr magneton.

For the partially spin-polarized incoherent phase, the single
layer free energy is
\be
\frac{E_i}2=\frac1{2\chi}M^2-MB_{tot},
\ee
where the magnetization
\be
M&=\frac12|g|\mu_B(n_{\uparrow}-n_{\downarrow})\equiv |g|\mu_B\Delta n,\\
\ee
and $\chi$ is the single layer spin susceptibility. The steady state is obtained
by minimizing $E_i$ with respect to $M$:
\be
\chi=\frac{M}{B_{tot}},
\ee
therefore
\be
\frac{E_i}2=\left\{\begin{array}{ll}-\frac12\chi
B_{tot}^2, & B_{tot}<B_{tot,p} \\ \frac1{2\chi}M_{max}^2-M_{max}B_{tot},&
B_{tot}>B_{tot,p} \end{array}\right.,
\ee
where the maximum magnetization $M_{max}$ and the field for full polarization $B_{tot,p}$ are given by
\be\label{Bp}
M_{max}&=\frac12|g|\mu_Bn=\frac{e|g|\mu_BB_{\perp}}{8\pi \hbar},\\
B_{tot,p}&=\frac{M_{max}}{\chi}.
\ee
Plugging these forms of free energy into (\ref{spin}), we obtain an equation
\be\label{spin2}
\frac{\d\delta_c}{\d B_{tot}}=\left\{\begin{array}{ll}\frac{-2\chi
B_{tot}+\frac{e|g|\mu_BB_{\perp}}{4\pi\hbar}}{\eta\frac{e^2}{\epsilon l^3}}, &
B_{tot}<B_{tot,p}  \\ 0 , & B_{tot}>B_{tot,p}  \end{array}\right..
\ee
Note that the RHS also depends on $\delta_c$ through $B_{\perp}$ which
determines $\ell$. Eqn.
(\ref{spin2}) can be solved numerically to yield the $\delta_c-B_{tot}$ curve.
For typical experimental parameters, $\d\delta_c/dB_{tot}$ starts out to be
positive when $B_{tot}$ is small, and continuously decreases to zero when
\be \label{Bp2}
-2\chi B_{tot}+\frac{e|g|\mu_BB_{\perp}}{4\pi\hbar} = 0,
\ee
this is nothing but Eqn. (\ref{Bp}) which determines the magnetic field at which
all composite fermions get polarized.

It remains to determine the
value of the composite fermion spin susceptibility $\chi$.
This can be done if $B_{tot}$ and $B_{\perp}$ at which full
polarization occurs are known, because from Eqn. (\ref{Bp}) or (\ref{Bp2}) we
have
\be\label{chi}
\chi=\frac{|g|\mu_BB_{\perp,p}}{4B_{tot,p}\phi_0},
\ee
where the subscript $p$ denotes the point of full polarization. In experimental
and exact-diagonalization studies,
one often parametrize $\chi$ with the form of non-interacting Fermi gas with a
``polarization mass'' $m_p$\cite{ParkJain1998,Shankar2001}:
\be
\chi=\frac{m_p}{4\pi\hbar^2}(|g|\mu_B)^2.
\ee
In the lowest-Landau-level approximation, $\frac{e^2}{\epsilon l}$ is the only relevant energy scale, and thus
\be
\frac{\hbar^2}{l^2m_p}\propto\frac{e^2}{\epsilon l}.
\ee
Therefore, presumably $m_p$ scales as $\sqrt{B_{\perp}}$:
\be
m_p=xm_e\sqrt{B_{\perp}},
\ee
where $m_e$ is the vacuum electron mass, $x$ is a dimensionless
number, $B_{\perp}$ is in units of Tesla. It is worth noting that
unlike free electrons spin-susceptibility which is proportional to
$1/m_e$, the susceptibility of composite fermions is proportional to
$m_p$ and therefore to $\sqrt{B}$. The reason for this is that the
Bohr magneton $\mu_B$ depends on the bare mass of the electron, and
therefore does not overturn the proportionality to effective mass in
the density of states factor of the susceptibility.

\begin{figure}
\centering
\includegraphics[scale=0.5]{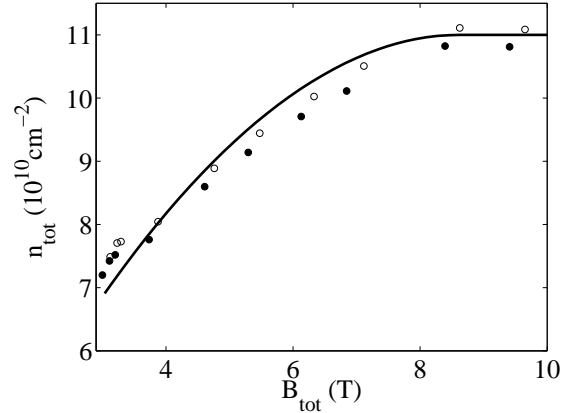}
\caption{Total electron density deduced from the critical $d/l=\delta_c$ vs. the total magnetic field for the parallel magnetic field experiments. Open and solid circles are experimental results of Giudici {\it et al.} \cite{Fujisawa2008} (c.f. FIG. 4a there). Solid line is our theoretical calculation with the fitting parameter $\eta=0.8\cdot10^{-3}$. The boundary condition in our calculation is chosen as $n_{tot}=11\cdot 10^{10}$cm$^{-2}$ when $B=10$T. }\label{parallel_field}
\end{figure}

For the parallel field experiment of Ref. \onlinecite{Fujisawa2008}, composite
fermions get polarized at total density $n_{tot}=11\cdot 10^{10}$cm$^{-2}$, tilting angle
$\theta=58^{\circ}$,  which corresponds to $B_{tot,p}=8.60$T,
$B_{\perp,p}=4.56$T, $x=0.56$ if we parametrize $\chi$ in terms of the
polarization mass $m_p$. Then we solve the ODE (\ref{spin2}) with the boundary condition at the high field endpoint ($B_{tot}=10$T, $n_{tot}=11\cdot 10^{10}$cm$^{-2}$), and plot the
$n_{tot}$ deduced from  $\delta_c$ vs. $B_{tot}$ in FIG
.\ref{parallel_field}. To tune the result to resemble the experimental results
in
FIG. 4a of Ref. \onlinecite{Fujisawa2008}, we get
\be
\eta=(0.8\pm0.2)\cdot10^{-3},
\ee
where the error mainly comes from fitting errors, meaning a finite range of
$\eta$'s make the $\delta_c-B_{tot}$ curve resemble the experimental result.

For the NMR and heat pulse experiments of Ref. \onlinecite{Spielman2005}, the
phase boundary before any perturbation is $\delta_{c0}=1.967$, which correspond
to $B_{\perp}=3.26$T. Ref. \onlinecite{Spielman2005} has estimated the effective
nuclear magnetic field to be $B_N=-0.17$T, therefore the total effective
magnetic field felt by electronic spin is $B_{tot}=B_{\perp}+B_N$. After a heat
pulse, nuclear spins are depolarized, and $B_N$ is set to zero. $B_{tot}$ is
strengthened to $B_{\perp}$, and the phase boundary changes to $\delta_c=1.983$.
We can not determine the spin susceptibility or the polarization mass
directly from experimental information, and therefore we use the numerical and experimental results from the
literature $m_p=(0.7\pm0.2)m_e\sqrt{B_{\perp}}$ with $B$ in units of Tesla
\cite{ParkJain1998,Melinte2000,Freytag2002,Kukushkin1999,Tracy2007}. In
this way, we obtain
\be
\eta\approx(1.3\pm0.4)\cdot10^{-3},
\ee
where the error mainly comes from uncertainty in the value of the polarization
mass $m_p$.

Note that our calculations in this section do not rely on the Chern-Simons description of composite fermions.

\section{Finite Temperature Transition Experiments}\label{sec:finiteT}

Ref. \onlinecite{Champagne2008} has studied the changes in critical $\delta_c=d/l$ as a function of the temperature $T$. They found that the phase boundary moves to smaller $d/l$ with higher $T$. When analyzing the temperature dependence of the transition, one needs to include the entropy contributions to the free energy associated with various low energy excitations for both phases. In the interlayer-coherent quantum Hall phase, the only gapless excitation is the linearly dispersing Goldstone mode, which corresponds to in-plane spin wave in the pseudospin language. Therefore, this mode dominates the temperature dependence of the free energy of the coherent phase (see \onlinecite{note1}). Denoting its velocity to be $v$, we have the free energy
\be
E_c(T)&=\sum_kT\ln(1-e^{-\hbar vk/T})\approx \frac{-1.2}{2\pi}\frac{T^3}{(\hbar v)^2},
\ee
and therefore
\be
\frac{\partial E_c}{\partial T}=-\frac{1.8}{\pi}\frac{T^2}{(\hbar v)^2}.
\ee
We use the experimental result of Ref. \onlinecite{Spielman2001} to
estimate the value of $v$ (which we assume to be a constant independent of $\delta$):
\be
v=1.4\cdot10^4 m\cdot s^{-1}
\ee

\begin{figure}
\includegraphics[scale=0.5]{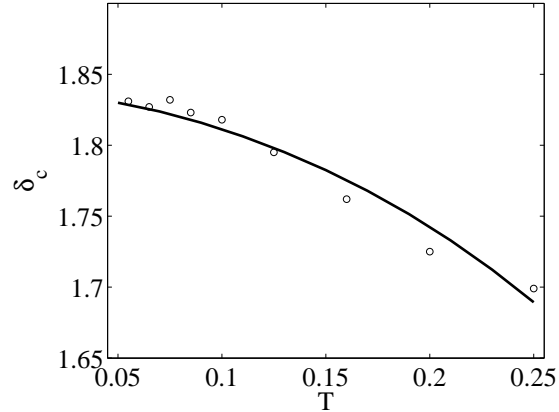}
\caption{The phase boundary $d/l$ vs. the temperature $T$ (in Kelvin) for the finite temperature experiments. Circles are experimental results of Champagne {\it et al.} \cite{Champagne2008} (c.f. FIG. 2c there). Solid line is our theoretical calculation with the fitting parameter $\eta=0.7\cdot10^{-3}$. The boundary condition in our calculation is chosen as $\delta_c=1.83$ when $T=50$mK. }\label{fig_finiteT}
\end{figure}

For the incoherent phase, working in the Chern-Simons framework, we have contributions from composite fermions as well as Chern-Simons gauge fields. The free energy is
\be
E_i&=-T\ln Z,\\
\ee
where the partition function $Z$ contains both composite fermion fields and
Chern-Simons gauge fields of the two layers. Integrating out the
composite fermions, we obtain\cite{HalperinLeeRead,Kim1996} (see Appendix \ref{appendix:finiteT} for details)
\be
Z=Z_0Z_{+}Z_-,
\ee
where $Z_0$ is the partition function for free fermions, and
\be
Z_{\pm}&=\int \mD a_{\pm}e^{-\int\d\tau\d^2x (a_{\pm}D^{-1}_{\pm}a_{\pm}/2)},
\ee
where $a_{\pm}$ are the in-phase and out-of-phase
combinations of Chern-Simons gauge fields of the two layers, and the polarizations $D_{\pm}^{-1}$ in the Coulomb gauge have the following form
\be
D^{-1}_{\pm}=\frac12\left(\begin{array}{cc} \Pi^0_{00} & \frac{iq}{4\pi} \\
\frac{-iq}{4\pi} & \Pi^0_{11}+\frac{2V_{\pm}q^2}{(4\pi)^2}  \end{array}\right),
\ee
where the index 0 and 1 denote time and transverse component, respectively.
\be
V_{\pm}(q)&=\frac12\left[\frac{2\pi e^2}q(1\pm e^{-qd})\right]F(q)
\ee
is linear combinations of intralayer and interlayer Coulomb interactions, $F(q)$ is the finite thickness form factor\cite{Price1984,Gold1987}, and $\Pi^0_{00}$ and $\Pi_{11}^0$ are the fermion density and transverse current correlations functions, respectively:
\be
\Pi^0_{00}&\approx\frac{m_*}{\pi}\left(1+i\frac{\omega}{v_Fq}\right),\\
\Pi^0_{11}&\approx-\frac{q^2}{12\pi m_*}+i\frac{2n\omega}{k_Fq}.
\ee
$m^*$ is the activation mass of the composite fermions, and, as we
discuss below is different from the polarization mass $m_p$ used in
the previous section. Continuing the derivation,
\be
E_i=-T\ln Z=-T\ln Z_0-T\ln Z_{+}-T\ln Z_-,
\ee
where the
free fermion part gives
\be\label{finiteT_fermion}
\frac{\partial E_{i,fermion}}{\partial T}=-\frac{\partial (T\ln Z_0)}{\partial T}=-\frac{2\pi}{3}T\frac{m_*}{\hbar^2},
\ee
and the gauge field parts give\cite{Wen,HalperinLeeRead}
\be
\frac{\partial E_{i,\pm}}{\partial T} =-\int_0^{\infty}\frac{\omega \d\omega}{\pi
T^2}\frac{e^{\beta\omega}}{(e^{\beta\omega}-1)^2}\int_0^{\infty}\frac{q\d q}{2\pi}
\im\ln\det D_{\pm}^{-1},
\ee
 A straightforward calculation following Ref. \onlinecite{HalperinLeeRead} shows that in the zero-thickness
approximation (form factor $F(q)$ set to 1),
\be
\frac{\partial E_{i,\pm}}{\partial T}=
-\frac{1.917}{4\pi}\cdot\frac53C_1^{2/3}T^{2/3}-\frac{1.645C_2}{2\pi^2}T\ln\frac{\omega_0}{T},
\ee
where
\be
C_1=\frac{16\pi n}{k_Fde^2/\epsilon}, C_2=\frac{8\pi n}{k_Fe^2/\epsilon},
\omega_0=\frac{(2k_F)^2}{C_2},\nonumber
\ee
$n$ is the single layer density of composite fermions, and $k_
F=\sqrt{2\pi n}$. 

Finite thickness corrections to the form of Coulomb interaction is found to have
negligible effect on the value of $\eta$, partly because it only affects the gauge
field contribution which is itself dominated by the free
composite-fermion-quasiparticle contribution for experimentally
relevant temperatures and for the choice of $m_*$ discussed below.

The value of the composite fermion mass $m_*$ is believed to be close to the
value determined by the activation gaps of fractional quantum Hall phases
away from $\nu=1/2$ \cite{HalperinLeeRead,SternHalperin1995,Kim1996,Simon1998}.
Therefore, we use the experimental value of this activation mass
determined from gap measurements in Refs. \onlinecite{Du1993,Du1994}, which is
\be
\frac{m_*}{m_e\sqrt{B_{\perp}}}=0.2\pm0.02.
\ee
Note that in numerical calculations the activation mass is typically smaller than experimental value by about a factor of 2\cite{HalperinLeeRead,Morf1995,ParkJain1999},  but it is believed that the theoretical value should approach experimental value once finite thickness effect, disorder and Landau level mixing are taken into account\cite{Yoshioka1986,ParkJain1999,Morf1999,ParkMeskiniJain1999}. Therefore, we feel the use of experimental value stated above is more appropriate. Also note that the polarization mass $m_p$ we used in the previous section is different from the mass we use here. Conceptually, within the Landau Fermi liquid theory, the two masses are related by $m_p=m_*/(1+F_0^a)$, $F_0^a$ being the zeroth spin-asymmetric Landau parameter.

Using this value of the mass along with the forms of free energy
in Clausius-Clapeyron equation (\ref{finiteT}) , we get an ODE, which can be solved with the boundary condition that $\delta_c=1.83$ when $T=50$mK to yield the
$\delta_c-T$ curve plotted in FIG. \ref {fig_finiteT}. To make this curve
resemble the
experimental result of Ref. \onlinecite{Champagne2008}, we have set
\be
\eta = (0.7\pm0.2)\cdot 10^{-3},
\ee
where the error mainly comes from the uncertainty in the value of the
activation mass $m_*$ and also the fitting error, meaning a finite range of
$\eta$'s make the $\delta_c-T$ curve resemble the experimental result.

In the above calculation, we assumed that the composite Fermi liquid is spin-unpolarized, and one might wonder how partial spin-polarization would affect the result. Because the free fermion contribution dominates $\partial E_i/\partial T$ and it is proportional to the density of states of composite fermions, our results would stay the same for partially-polarized composite Fermi liquid.

\section{Density Imbalance Experiments}\label{sec:imbalance}

Refs. \onlinecite{Spielman2004,Champagne2008b} have studied the dependence of the critical $\delta_c=d/l$ on the density imbalance between the layers. They observed that at small imbalance, the phase boundary has a quadratic dependence on the density imbalance, and the coherent quantum Hall phase survives at higher $d/l$ with larger imbalance.

Denoting the density of the two layers $n_{1,2}$, a density imbalance between the layers,
\be
\Delta n\equiv\frac{n_1-n_2}2
\ee
costs an energy which includes a dominating geometrical capacitance term and quantum mechanical corrections. This is true for both phases.
For the coherent phase, we follow Ref. \onlinecite{Moon1995} to obtain the free energy density to be
\be
E_c&=\left(\frac{2\pi e^2d}{\epsilon}+\beta_{m,E}\right)(\Delta n)^2,\\
\beta_{m,E}&=\int_0^{\infty}\frac{qdq}{2\pi}V^z(q)h(q)
\ee
where $2\pi e^2d/\epsilon$ is the geometrical capacitance term, while $\beta_{m,E}$ is the exchange contribution which tends to offset the geometrical capacitance term. Here, $V^z(q)=V(q)-U(q)$, $V(q)=\frac{2\pi e^2}{\epsilon q}F(q)$ is the intralayer
Coulomb interaction,  $F(q)$ is the finite thickness form factor\cite{Price1984,Gold1987},
$U(q)=V(q)e^{-qd}$ is the interlayer Coulomb interaction, and $h(q)=-2\pi
l^2\exp(-{q^2l^2}/{2})$ is the pair distribution function of the Halperin (1,1,1)
wavefunction.

The free energy density of the incoherent phase is (see Appendix \ref{appendix:imbalance} for details)
\be\label{cost}
E_i
=\frac{(\Delta n)^2}{\tilde{K}-\tilde{K}'},
\ee
where
\be
\tilde{K}&\equiv \frac1{\beta A}\lim_{\q\rightarrow0}\lim_{\o\rightarrow0}\la\rho_{1,\q,\o}\rho_{1,-\q,-\o}\ra,\\
\tilde{K}'&\equiv \frac1{\beta A}\lim_{\q\rightarrow0}\lim_{\o\rightarrow0}\la\rho_{1,\q,\o}\rho_{2,-\q,-\o}\ra,
\ee
where $\beta$ is the inverse of temperature, $A$ is the area of the sample, $\rho_{1,2}$ are the composite fermion density of each layer.
Treating the Coulomb interaction within RPA, we obtain (see Appendix \ref{appendix:imbalance} for details)
\be\label{K_result}
\tilde{K}&=-\tilde{K}'=\frac{\kappa}{2\left(1+\frac{2\pi e^2d}{\e}\cdot \kappa\right)},
\ee
where $\kappa$ is the $\omega\rightarrow0,q\rightarrow0$ limit of the 1-particle-irreducible density response function, namely compressibility, of a single-layer composite Fermi liquid. Plugging (\ref{K_result}) into 
(\ref{cost}), one obtains the energy cost of uniform density imbalance in the incoherent phase:
\be
E_i&=\left(\frac1{\kappa}+\frac{2\pi e^2d}{\e}\right)\Delta n^2.
\ee
From the Clausius-Clapeyron equation (\ref{imbalance}), the geometrical capacitance term of the two phases cancels
out, and we have
\be \label{imbalance2}
\eta=\frac{\kappa^{-1}-\beta_{m,E}}{\frac{\d\delta_c}{d (\Delta
n^2)}\frac{e^2}{\epsilon l^3}}.
\ee

Since $\kappa$ is the single layer compressibility, it is connected to the ground state energy per area of the composite Fermi liquid $E_{GS}$ via
\be\label{P_GS}
\kappa^{-1}=\frac{\partial^2E_{GS}}{\partial n^2}.
\ee
Note that our definition of the compressibility is slightly different from some literature where $\kappa^{-1}=n^2\frac{\partial^2E_{GS}}{\partial n^2}$ are used instead.

Alternatively, treating the Chern-Simons interaction within RPA (see Appendix \ref{appendix:imbalance} for details), we obtain
\be\label{P_result_CS}
\kappa^{-1}&=\kappa_0^{-1}-{16\pi^2}\chi_{d},
\ee
where
\be
\kappa_0=\frac{m_*}{\pi(1+F_0^s)}
\ee
is the compressibility without the Chern-Simons interaction, $F_0^s$ is the zeroth Landau parameter in the spin-symmetric channel, and
\be
\chi_d=-\frac{1}{12\pi m_*}
\ee
is the Landau diamagnetic susceptibility.
Therefore
\be\label{kappa_cs}
E_i
&=\left(\frac{\pi}{m_*}+\frac{\pi F_0^s}{m_*}+\frac{4\pi}{3m_*}+\frac{2\pi e^2d}{\e}\right)\Delta n^2.
\ee
Clearly, we can identify the four terms as free fermion contribution, exchange/correlation effect, Landau diamagnetism for Chern-Simons flux\cite{Champagne2008b}, and geometric capacitance term, respectively.

Although the Chern-Simons expression of $\kappa$ Eqn. (\ref{kappa_cs}) offers valuable physical insight into its structure, the precise value of the parameters $m_*$,
$\chi_d$, and especially $F_0^s$ are not very well understood. The best way to estimate $\kappa$ is to use its connection with ground state energy density $E_{GS}$ of composite Fermi liquid (\ref{P_GS}).
In the zero-thickness approximation, Park {\it et
al.} \cite{Park1998} have estimated the value of $E_{GS}$ for spin unpolarized composite Fermi liquid to be
\be
E_{GS}=-0.4695\frac{e^2}{\epsilon l}n,
\ee
thus
\be\label{kappa1}
\kappa^{-1}=-0.4695\cdot3\pi\frac{e^2}{\epsilon}l,
\ee
where $n$ is the single layer density of composite fermions, and $l$ is the magnetic length.

\begin{figure}{
\includegraphics[scale=0.5]{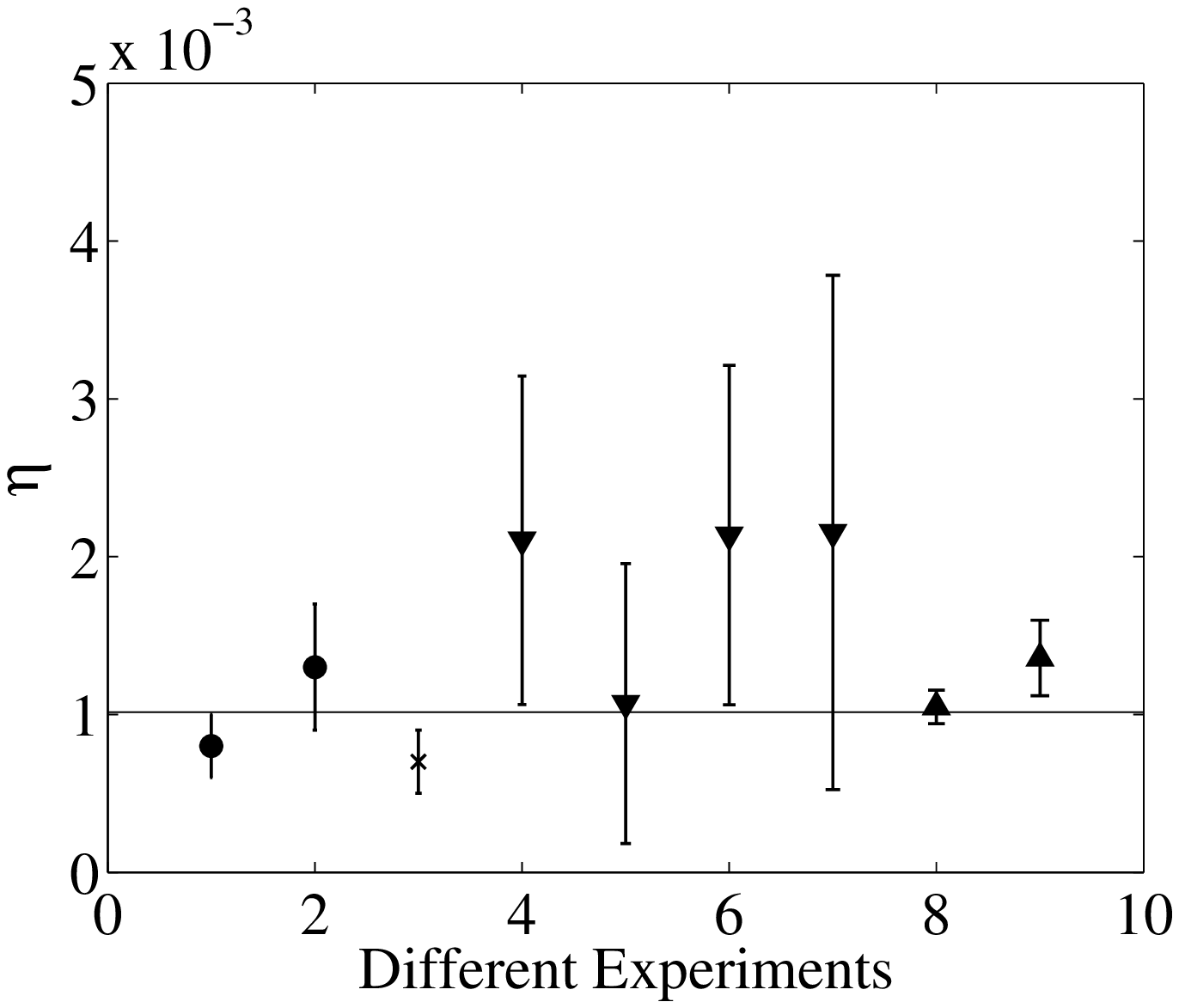}
}
\caption{Summary of the value of $\eta$ extracted from various experiments. Experiment 1: parallel field experiment of Ref. \onlinecite{Fujisawa2008}.
Experiment 2: NMR/heat pulse experiment of Ref. \onlinecite{Spielman2005}.
Experiment 3: finite temperature transition experiment of Ref. \onlinecite{Champagne2008}. Experiment 4 to 7: density imbalance experiments of Ref. \onlinecite{Champagne2008b}, with T = 55mK, 85mK, 125mK, 200mK.
Experiment 8 and 9: density imbalance experiments of Ref.
\onlinecite{Spielman2004} with phase boundary determined by Hall drag and
tunneling.
To obtain this result we used the numerical result of Ref. \onlinecite{Park1998} for unpolarized
composite Fermi liquid ground state energy to estimate $\kappa^{-1}$. The horizontal line is the average value of $\eta$ weighted by inverse of error square, which is $\sim(1\pm0.1)\times10^{-3}$.
}\label{summary}
\end{figure}

Using this value of $\kappa^{-1}$ and the zero-thickness form of Coulomb
interaction to calculate the coherent phase exchange term $\beta_{m,E}$
(because the numerical result for $E_{GS}$ of the incoherent phase quoted above
from Ref. \onlinecite{Park1998} was also done with zero thickness), and extracting
the curvature $\frac{\d\delta_c}{d (\Delta n^2)}$ from experiments, we readily
obtain the value of $\eta$. This result does not depend on the Chern-Simons description of composite fermions. We have plotted in FIG. \ref{summary} the values of $\eta$ extracted
from density imbalance experiments as well as those determined from spin transition and finite temperature
transition experiments. The error
bars for the density imbalance experiments mainly come from fitting errors.

Note that the main effect of the finite thickness correction
to the form of Coulomb interaction is to reduce the exchange terms of both
phases. Since the value of $\eta$ is related to the difference between the exchange
term of the two phases, we do not expect the result of $\eta$ to sensitively
depend on this effect. Nevertheless, we can include it in the Chern-Simons treatment of $\kappa$. We use the activation mass $m_*=0.2m_e\sqrt{B_{\perp}}$ estimated in Sec. \ref{sec:finiteT} as the value of $m_*$, set $\chi_d=-1/(12\pi m_*)$, and use the Hubbard approximation to estimate $F_0^s$. In the
Hubbard approximation, the exchange effect is taken into account by introducing
a many-body local field factor $G(q)
={q}/(2\sqrt{q^2+k_F^2}),
$
and $F_0^s=-\frac{m_*}{\pi}\lim_{q\rightarrow0}V(q)G(q)$. Thus, we obtain from Eqn. (\ref{P_result_CS})
\be\label{kappa2}
\kappa^{-1}=\frac73\frac{\pi\hbar^2}{m_*} - \frac{\pi e^2}{\epsilon k_F}.
\ee

Using this value of $\kappa^{-1}$ and the
finite-thickness form of Coulomb interaction to calculate the coherent phase
exchange term $\beta_{m,E}$, we have calculated the values of
$\eta$ from density imbalance experiments which turned out to be extremely close to the results obtained earlier in FIG. \ref{summary}.

Comments about the value of the compressibility in the composite Fermi liquid phase are in order. First, In Ref. \onlinecite{Eisenstein1994}, the compressibility of a single layer 2DEG at zero field was studied in detail, and it was found that aside from the well-known density-of-states contribution and exchange contribution to the compressibility, there is a third contribution coming from the so-called Hartree band-bending effect due to the influence of the finite quantum well width on the out-of-plane direction of electron wavefunction. For the bilayer system studied here, we expect a similar effect on the composite Fermi liquid compressibility $\kappa^{-1}$ in the incoherent phase and on $\beta_{m,E}$ for the coherent quantum Hall phase as well. A quantitative analysis of this effect and its impact on the density imbalance experiments is beyond the scope of this paper, and we simply note that the Hartree band-bending effect is essentially a single-particle effect\cite{Eisenstein1994}, and therefore it will contribute equally to $\kappa^{-1}$ and $\beta_{m,E}$. To obtain the value of $\eta$ from Eqn. (\ref{imbalance2}), we only need the difference between $\kappa^{-1}$ and $\beta_{m,E}$, and therefore we do not expect the Hartree band-bending effect to modify our results. Second, quenched disorder acts to broaden the Landau levels and therefore adds a positive contribution to the compressibility. This could account for the close-to-zero compressibility measured by Ref. \onlinecite{Eisenstein1994}. Again, this effect is likely to be similar for both phases, and we do not expect disorder to affect the difference between $\kappa^{-1}$ and $\beta_{m,E}$ appreciably. Nevertheless, disorder is important in smearing the first order transition into a continuous one (see discussion in Sec. \ref{sec:summary}).

We assumed that the composite Fermi liquid is unpolarized above, but again we do not expect partial polarization to affect our results strongly. For (\ref{kappa1}), Park {\it et al.}\cite{Park1998} also reported the ground state energy for polarized composite Fermi liquid to be very close to the unpolarized one quoted above:
\be 
E_{polarized}=-0.4656\frac{e^2}{\epsilon l}n,
\ee
and therefore our results would also stay very close. In the Chern-Simons treatment (\ref{kappa_cs}) and(\ref{kappa2}), since the Chern-Simons fields couple to both spins and the density and current response function stays the same for partially-polarized and unpolarized composite Fermi liquids, our calculation also remains valid (see Appendix \ref{appendix:imbalance}).

\section{Summary And Discussion}\label{sec:summary}

To summarize, we derived the Clausius-Clapeyron relations
[Eqn. (\ref{spin}, \ref{finiteT}, \ref{imbalance})] for the phase
transition tuned by $d/l$ in bilayer $\nu_{tot}=1$ quantum Hall
system, assuming that it is a first-order transition between
spin-polarized coherent quantum Hall state and spin
partially-polarized composite-fermion Fermi liquid state. In
Sec. \ref{sec:magnetic}, we studied the changes of phase boundary
$(d/l)_c$ when the magnetic field coupled to spin is changed by either
NMR/heat pulse or parallel magnetic field. The phase boundary as a
function of temperature was studied in Sec. \ref{sec:finiteT}. The
temperature dependence of free energy in the coherent quantum Hall
phase is dominated by the linearly-dispersing Goldstone mode, while
the incoherent composite Fermi liquid phase has contributions from
both fermions and gauge fields. In Sec. \ref{sec:imbalance}, we
investigated the changes of phase boundary when there is density
imbalance between the two layers. We use the result of
Ref. \onlinecite{Moon1995} for the free energy cost of density
imbalance in the coherent quantum Hall phase. The free energy for the
incoherent phase is shown to be connected to the compressibility of
single layer composite Fermi liquid.

Our main goal was to check the consistency of the Clausius-Clapeyron
relation with the observed transition. Each experiment which observes
the change in $(d/\ell)_c$ due to changing another parameter in the
system indicates a value for $\eta$, as defined in Eq. (\ref{eta});
all values should agree. 

In FIG. \ref{summary}, we have plotted the values of $\eta$ determined from spin transition, finite temperature transition, and density imbalance transition experiments. The horizontal line is the average value of $\eta$ weighted by inverse of error square, i.e., the maximum likelihood estimator of $\eta$. 
One can see that, indeed, all nine values of $\eta$ extracted
from various experiments roughly lie in the range
$1\sim2\times10^{-3}$, and the weighted average value of $\eta=(1\pm0.1)\cdot10^{-3}$ is roughly within all the error
bars. Our analysis, therefore, confirms the consistency for the
scenario of a direct first-order phase transition between coherent
quantum-Hall phase and incoherent composite Fermi-liquid
phase. Furthermore, the analysis provides a unified framework within
which we can understand the observed phase boundaries for several
distinct experiments.

\begin{figure}
\includegraphics[scale=0.35]{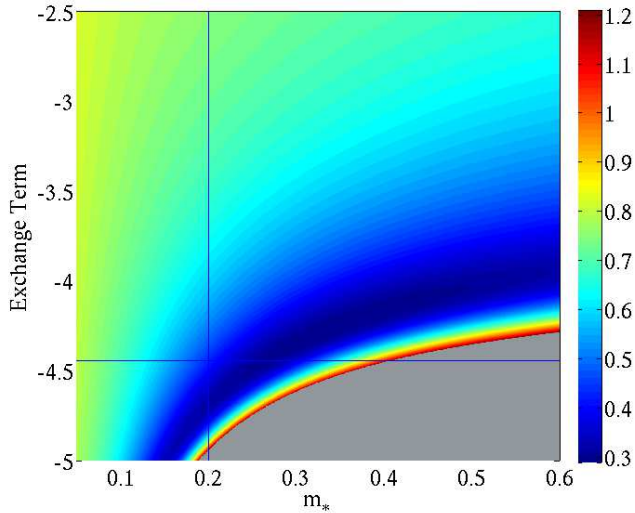}
\caption{(Color online.) Standard deviation of $\eta$ among various experiments
divided by their average value (which measures the goodness of agreement between $\eta$s) within the Chern-Simons framework as a function of
the composite fermion mass and the exchange contribution to $\kappa^{-1}$ [see Eq. (\ref{kappa2})].
Horizontal axis: composite fermion mass in units of $m_e\sqrt{B_{\perp}}$,
$m_e$ being the vacuum electron mass, $B_{\perp}$ is in units of Tesla. Vertical axis: exchange contribution to
$\kappa^{-1}$, which is  $F_0^s\pi/m_*$ (in units of $e^2l/\epsilon$, $l$ being the magnetic length). Grey color denotes the region where at least one of the $\eta$'s becomes negative, thus
unphysical. The horizontal line denotes the Hubbard approximation to the
exchange effect ($-\sqrt2\pi$). The vertical line denotes the experimental value of the activation mass
$m_*\approx0.2m_e\sqrt{B_{\perp}}$, which is the value of
composite fermion mass we used in calculations for FIG. \ref{fig_finiteT} and
FIG. \ref{summary}.}\label{sweep}
\end{figure}

In Sec. \ref{sec:imbalance}, we also worked in the Chern-Simons description of composite fermions [i.e. Eq. (\ref{kappa2})] in addition to our treatment [i.e. Eq. (\ref{kappa1})] using the numerical results of Ref. \onlinecite{Park1998}, and we obtained extremely similar results. Stepping back a little from that analysis with the Chern-Simons treatment, one can pretend ignorance of any knowledge of the parameters including the effective
mass $m_*$ and the exchange contribution to $\kappa^{-1}$, and ask what values of them would give good agreement between the values of $\eta$ extracted from experiments. We
have plotted the standard deviation of $\eta$ extracted from various
experiments divided by the their average value in
FIG. \ref{sweep} as a function of the composite fermion mass (in units of $m_e\sqrt{B_{\perp}}$) and the exchange
contribution to $\kappa^{-1}$, which is $F_0^s\pi/m_*$ (in units of $e^2l/\epsilon$). The finite thickness form of the Coulomb interaction
is used in calculating the coherent-phase exchange term  when producing
this plot. Grey color denotes the region where at least one of the $\eta$'s becomes negative, thus
unphysical, while dark blue denotes parameter regimes which give rise to good agreement among $\eta$'s extracted from different experiments. The horizontal line denotes the Hubbard approximation to the
exchange effect ($\sqrt2\pi$), while he vertical line denotes the experimental value of the activation mass
$m_*\approx0.2m_e\sqrt{B_{\perp}}$.

We have not explicitly discuss the role of disorder, which is always present in the samples. Disorder will bring spatial fluctuations into some variables in the Clausius-Clapeyron equations we have derived, and therefore smear the first order transition into a continuous one, as observed in experiments. Roughly speaking, the analysis we have performed in this work applies to the spatially averaged quantities. For example, with disorder, the RHS of the Clausius-Clapeyron equation for spin transitions (\ref{spin2}) will acquire spatial dependence most likely through a spatially fluctuating spin susceptibility $\chi$:
\be\label{spin3}
\frac{\d\delta_c(\vec x)}{\d B_{tot}}=\frac{-2\chi(\vec x)
B_{tot}+\frac{eg\mu_BB_{\perp}}{4\pi\hbar}}{\eta\frac{e^2}{\epsilon l^3}}
\ee
Thus, one can take the spatial average of both sides and study how the averaged critical $\delta_c$ changes with $B_{tot}$, as we did in this work. Furthermore, one can also take the standard deviation of both sides of (\ref{spin3}), and conclude that the width of the phase transition, which is the standard deviation of $\delta_c$, grows with $B_{tot}$ assuming the standard deviation of $\chi(\vec x)$ does not change appreciably with $B_{tot}$. One can also study the finite temperature transition in a similar way. Because there the free fermion term (\ref{finiteT_fermion}) dominates, one can conclude that the transition becomes wider at higher temperature, if one assumes the composite fermion mass $m_*$ has some temperature-independent spatial variation. This is in accord with the experimental observation of Ref. \onlinecite{Champagne2008}.

A major question which is not directly addressed in our analysis is the
possibility of a continuous quantum crossover between the coherent
and incoherent phases (see, e.g.,
Refs. \onlinecite{Moller2008a,Moller2008}). If indeed no
real thermodynamic singularity exists even in the clean case, then there is
no reason for the Clausius-Clapeyron relations to hold as well as we
find they do. Nonetheless, there is also no contradiction in them
holding where no first-order transition exists. In this case, however, we can draw the conclusion that the
crossover region between the two phases must be very narrow, such
that it approximates a smeared thermodynamic singularity (just as disorder would
widen a thermodynamic singularity) and therefore follows the
Clausius-Clapeyron relations we presented here for the unmixed
phases. In other words, a good agreement with the relations indicates
that already at regions in parameter space close to the transition,
the thermodynamic functions of the pure coherent
and pure incoherent phases apply, and they indicate a smeared phase
transition line.

Additional outstanding questions which we did not address, but are
noteworthy are as follows. First, a thermodynamic
phase transition between the coherent and incoherent phases does not
have to be first order at high Zeeman fields when both phases are
spin-polarized; a second-order phase transition is not ruled out {\it
  a priori}. Future experiments should clarify this
issue (see the recent experiments of Refs. \onlinecite{Finck2010,Muraki2010}). In addition, for the density imbalance transitions, we have
mainly focused on the regime of small imbalance, while the experiments
of Ref. \onlinecite{Champagne2008b} have studied the case of
large imbalance, e.g., $\Delta\nu=\nu_1-\nu_2\leq0.4$. The interlayer
incoherent phase in that regime could be two decoupled single-layer
fractional quantum Hall phase. It would be very interesting to see if
a similar Clausius-Clapeyron equation can describe the phase
transition in that case. Finally, although our assumption (\ref{eta})
is very natural on qualitative ground, a
microscopic derivation of this quantity would be very useful.

\acknowledgments
It is a pleasure to acknowledge useful conversations with J. Alicea, A.
Champagne, H. Fertig, A.D.K Finck, A.H. MacDonald, G. Murthy, F. von Oppen, E. Rezayi, S. Simon, and
I. Spielman. We thank K. Muraki for providing us the experimental data shown in FIG. \ref{parallel_field}. We are grateful for support from the research corporation,
the Packard Foundation, and the Sloan foundation (GR),the
Israel-US Binational Science Fundation, the Minerva Foundation, Microsoft
Station Q (AS),
and NSF grant DMR-0552270 (JPE).

\appendix

\section{DERIVATION OF THE TEMPERATURE DEPENDENCE OF THE COMPOSITE FERMI LIQUID FREE ENERGY}\label{appendix:finiteT}

Within the Chern-Simons description of the composite-fermion Fermi liquid at $\nu=1/2$, we have the following partition function of the system:
\be\label{action}
Z=\int \mD a_1\mD a_2\mD\psi_{1\s}\mD\psi_{2\s}e^{-\int\d\tau d^2x \mathcal L},
\ee
where
\be
\mathcal L&=\sum_{n=1,2}\left\{\psi^{\dagger}_{n\sigma}(\partial_{\tau}-\mu_n-ia_{n,0})\psi_{n\s}-\frac{i}{8\pi}a_{n\mu}\epsilon^{\mu\nu\lambda}\partial_{\nu}a_{n\lambda}\right.\\
&+\frac1{2m}\psi^{\dagger}_{n\s}\left(-i\nabla-\vec a_n\right)^2\psi_{n\s}\\
&\left.+\frac12\int \d^2x'\psi^{\dagger}_{n\sigma}(x)\psi_{n\sigma}(x)V(x-x')\psi^{\dagger}_{n\sigma'}(x')\psi_{n\sigma'}(x')\right\}\\
&+\int \d^2x'\psi^{\dagger}_{1\sigma}(x)\psi_{1\sigma}(x)U(x-x')\psi^{\dagger}_{2\sigma'}(x')\psi_{2\sigma'}(x'),
\ee
where $\psi_{n\s}$ is the composite fermion fields in the $n$'th layer with spin $\s$, $V$ and $U$ are the intralayer and interlayer Coulomb interaction, respectively. Here, $a_{n\mu}$ are the fluctuations of the Chern-Simons gauge fields in the $n$'th layer from its saddle point value which cancels the external magnetic field exactly, and $\mu=0,1,2$ are the time and two spatial coordinates, respectively. Integrating out $a_{n,0}$, one obtains the expected constraints
\be
\nabla\times\vec a_{n} = 4\pi\psi^{\dagger}_{n\s}\psi_{n\s}.
\ee
Following Ref. \onlinecite{HalperinLeeRead}, we make use of this constraint and replace $\psi^{\dagger}_{n\s}\psi_{n\s}$ in Coulomb interaction terms by $\nabla\times\vec a_{n}/(4\pi)$.
Next, we define
\be
a_{\pm\mu}&=a_{1\mu}\pm a_{2\mu},\\
V_{\pm} &=\frac{V\pm U}2,
\ee
and reorganize $\mathcal L$ as
\be
\mathcal L &= \mathcal L_f +\mathcal L_{CS},\\
\mathcal L_f&=\psi^{\dagger}_{1\sigma}\left(\partial_{\tau}-\mu_1-i\frac{a_{+0}+a_{-0}}2\right)\psi_{1\s}\\
&+\psi^{\dagger}_{2\sigma}\left(\partial_{\tau}-\mu_2-i\frac{a_{+0}-a_{-0}}2\right)\psi_{2\s}\\
&+\psi^{\dagger}_{1\s}\frac{\left(-i\nabla-(\vec a_++\vec a_-)/{2}\right)^2}{2m}\psi_{1\s}\\
&+\psi^{\dagger}_{2\s}\frac{\left(-i\nabla-(\vec a_+-\vec a_-)/2\right)^2}{2m}\psi_{2\s}\\
\mathcal L_{CS}&=-\frac{i}{16\pi}a_{+\mu}\epsilon^{\mu\nu\lambda}\partial_{\nu}a_{+\lambda}
-\frac{i}{16\pi}a_{-\mu}\epsilon^{\mu\nu\lambda}\partial_{\nu}a_{-\lambda}\\
&+\frac12\frac1{(4\pi)^2}\int \d^2x'[\nabla\times\vec a_{+}(x)]V_+(x-x')[\nabla\times\vec a_{+}(x')]\\
&+\frac12\frac1{(4\pi)^2}\int \d^2x'[\nabla\times\vec a_{-}(x)]V_-(x-x')[\nabla\times\vec a_{-}(x')].
\ee
Denoting the free fermion partition function to be
\be
Z_0=\int \mD\psi_{1\s}\mD\psi_{2\s}\exp\left(-\int\d\tau\d^2x \mathcal L_f(a_{\pm}=0)\right),
\ee
and following standard methods\cite{HalperinLeeRead} to integrate out
composite fermion fields $\psi_{n\s}$, we obtain
\be
Z=Z_0Z_{+}Z_-,
\ee
where $Z_0$ is the partition function for free fermions, and
\be
Z_{\pm}&=\int \mD a_{\pm}e^{-\int\d\tau\d^2x (a_{\pm}D^{-1}_{\pm}a_{\pm}/2)},
\ee
In Coulomb gauge, one can treat the polarizations $D^{-1}_{\pm}$ as $2\times2$ matrices, with index 0 and 1 to be the time and transverse component, respectively. Thus, $D^{-1}_{\pm}$ take the following form:
\be
D^{-1}_{\pm}=\frac12\left(\begin{array}{cc} \Pi^0_{00} & \frac{iq}{4\pi} \\
\frac{-iq}{4\pi} & \Pi^0_{11}+\frac{2V_{\pm}q^2}{(4\pi)^2}  \end{array}\right),
\ee
where $\Pi^0_{00}$ and $\Pi^0_{11}$ are the density and transverse current correlation functions of free fermions resulted from integrating out composite fermion fields.
Thus, the free energy is given by
\be
E_i=-T\ln Z=-T\ln Z_0-T\ln Z_{+}-T\ln Z_-,
\ee
and the rest of the steps are given in Section \ref{sec:finiteT}.

\section{DERIVATION OF THE FREE ENERGY FOR DENSITY IMBALANCE IN COMPOSITE FERMI LIQUID PHASE}\label{appendix:imbalance}

Starting from action (\ref{action}) or any other action for composite fermions, we integrate out all fluctuating fields and obtain
\be
&Z=\exp\left\{\frac1{2\beta A}\sum_{\vec q,\o}\left[K_{\q,\o}\phi_{1,\q,\o}\phi_{1,-\q,-\o}\right.\right.\\
&\left.\left.+K_{\q,\o}\phi_{2,\q,\o}\phi_{2,-\q,-\o}+2K'_{\q,\o}\phi_{1,\q,\o}\phi_{2,-\q,-\o}\right]\right\},
\ee
where
\be
K_{\q,\o}&=\frac1{\beta A}\la\rho_{1,\q,\o}\rho_{1,-\q,-\o}\ra=\frac1{\beta A}\la\rho_{2,\q,\o}\rho_{2,-\q,-\o}\ra,\\
K'_{\q,\o}&=\frac1{\beta A}\la\rho_{1,\q,\o}\rho_{2,-\q,-\o}\ra,
\ee
$\rho_j$ is the composite fermion density of the $j$'th layer, $\beta$ is the inverse of the temperature, $A$ is the area of the sample, and $\phi_{j,\q,\o}$ is the Fourier-transformed potential in the $j$'th layer.
For a constant potential $\phi_{j}$ ($j=1,2$), we have
\be
\phi_{j,\q,\o}=\phi_j\cdot\beta A\delta_{\q,0}\delta_{\o,0},
\ee
and the grand potential $\Omega$ is
\be
\Omega=-T\ln Z=-\frac{A}2\left(\tilde{K}\phi_{1}^2+\tilde{K}\phi_{2}^2+2\tilde{K}'\phi_{1}\phi_2\right),
\ee
where
\be
\tilde{K}\equiv \lim_{\q\rightarrow0}\lim_{\o\rightarrow0}K_{\q,\o},
\tilde{K'}\equiv \lim_{\q\rightarrow0}\lim_{\o\rightarrow0}K'_{\q,\o}.
\ee
The density in each layer is
\be
n_1&=-\frac1A\frac{\partial \Omega}{\partial \phi_1}=\tilde{K}\phi_1+\tilde{K}'\phi_2,\\
n_2&=-\frac1A\frac{\partial \Omega}{\partial \phi_2}=\tilde{K}\phi_2+\tilde{K}'\phi_1.
\ee
Finally, the free energy is obtained via a Legendre transformation
\be
F&=\Omega+\phi_1n_1A+\phi_2n_2A\\
&=\frac A4\left(\frac{(n_1-n_2)^2}{\tilde{K}-\tilde{K}'}+\frac{(n_1+n_2)^2}{\tilde{K}+\tilde{K}'}\right).
\ee
Within the RPA treatment of the Coulomb interaction, the full density response function $K$ is related to its one-particle-irreducible (1PI) counterpart ${\Pi}$ (which neglects the long range Coulomb interaction) by
\be\label{RPA}
K^{-1}={\Pi}^{-1}+\tilde V,
\ee
where $K$, ${\Pi}$, and $\tilde V$ are $2\times2$ matrices in the layer-index space:
\be
\tilde V=\left(\begin{array}{cc} V & U \\ U & V\end{array}\right),\qquad {\Pi}=\left(\begin{array}{cc} {\Pi}_{00} & 0 \\ 0 & {\Pi}_{00} \end{array}\right).
\ee
Here, $V$ and $U$ are intralayer and interlayer Coulomb interaction potential, respectively, and ${\Pi}_{00}$ in the static uniform limit gives the single layer compressibility $\kappa$:
\be
\kappa\equiv \lim_{q\rightarrow0}\lim_{\omega\rightarrow0}{\Pi}_{00}.
\ee
Solving (\ref{RPA}), we have
\be
K_{11}&=K_{22}=\frac{{\Pi}_{00}(1+{\Pi}_{00}V)}{(1+{\Pi}_{00}V)^2-{\Pi}_{00}^2U^2},\nonumber\\
K_{12}&=\frac{-({\Pi}_{00})^2U}{(1+{\Pi}_{00}V)^2-{\Pi}_{00}^2U^2}.
\ee
Given the form of Coulomb interactions
\be
V(q)=\frac{2\pi e^2}q F(q),\qquad U(q)=V(q)e^{-qd},
\ee
and the fact that the finite thickness form factor $F(q)\rightarrow1$ as $q\rightarrow0$,
in the limit $\omega\rightarrow0$ and $q\rightarrow0$, the denominators of $K_{11}$, $K_{22}$, and $K_{12}$ become
\be
&(1+{\Pi}_{00}V)^2-({\Pi}_{00})^2U^2\\
&\rightarrow\frac{4\pi e^2\kappa}{\e q}\left(1+\frac{2\pi e^2\kappa d}{\e}\right),\textrm{ as $\omega\rightarrow0,q\rightarrow0.$}
\ee
Therefore in this limit
\be
\tilde{K}&\equiv\lim_{q\rightarrow0}\lim_{\o\rightarrow0}K_{11}=\frac{\kappa}{2\left(1+{2\pi e^2\kappa d}/{\e}\right)},\\
\tilde{K}'&\equiv\lim_{q\rightarrow0}\lim_{\o\rightarrow0}K_{12}=-\frac{\kappa}{2\left(1+{2\pi e^2\kappa d}/{\e}\right)},
\ee
and the imbalance part of the free energy density is
\be
E_i&=\lim_{\q\rightarrow0}\lim_{\o\rightarrow0}\frac{\Delta n^2}{\tilde{K}-\tilde{K}'}\\
&=\left(\frac1{\kappa}+\frac{2\pi e^2d}{\e}\right)\Delta n^2,
\ee
as shown in Section \ref{sec:imbalance}. This result does not depend on the Chern-Simons description of composite fermions. Note also that the total compressibility $\tilde{K}+\tilde{K}'$ vanishes linearly in $q$ as $q\rightarrow0$ due to the long-range nature of the Coulomb interaction, similar to the single layer case as analyzed by Halperin {\it et al.} \cite{HalperinLeeRead}.

To calculate the single layer compressibility $\kappa$ within the Chern-Simons framework, we have the following RPA equation:
\be\label{CS_RPA}
({\Pi})^{-1}=(\Pi^0)^{-1}+C,
\ee
where $C$ is the propagator of the Cherns-Simons field, and $\Pi^0$ is the correlation functions without the Chern-Simons interaction. We work in the Coulomb gauge and treat ${\Pi}$, $\Pi^0$, and $C$ as $2\times2$ matrices in the space of density and transverse current. In the static and long wavelength limit, we have
\be\label{app_pi}
\Pi^0=\left(\begin{array}{cc} \kappa_0 & 0 \\ 0 & \chi_d q^2\end{array}\right),\qquad
C=\frac{4\pi}q\left(
\begin{array}{cc}
0 & i \\ -i & 0
\end{array}\right).
\ee
where $\kappa_0=m_*/[\pi\hbar^2(1+F_0^s)$] is the density response function neglecting Chern-Simons interaction, and $\chi_{d}$ is the Landau diamagnetic susceptibility. Hence,
\be
\kappa^{-1}=\kappa_0^{-1}-{16\pi^2}\chi_{d},
\ee
as shown in Section. \ref{sec:imbalance}. Note that these results are the same for unpolarized and partially-polarized composite Fermi liquids, because (\ref{CS_RPA}) is valid in any case since Chern-Simons fields couple to both spins, and the value of $\kappa_0$ and $\chi_d$ in (\ref{app_pi}) stays the same for partially-polarized composite Fermi liquid. The value of $F_0^s$ in the Hubbard approximation treatment is also roughly the same for partially-polarized and unpolarized composite Fermi liquids.

\bibliography{reference}

\begin{thebibliography}{68}
\expandafter\ifx\csname natexlab\endcsname\relax\def\natexlab#1{#1}\fi
\expandafter\ifx\csname bibnamefont\endcsname\relax
  \def\bibnamefont#1{#1}\fi
\expandafter\ifx\csname bibfnamefont\endcsname\relax
  \def\bibfnamefont#1{#1}\fi
\expandafter\ifx\csname citenamefont\endcsname\relax
  \def\citenamefont#1{#1}\fi
\expandafter\ifx\csname url\endcsname\relax
  \def\url#1{\texttt{#1}}\fi
\expandafter\ifx\csname urlprefix\endcsname\relax\def\urlprefix{URL }\fi
\providecommand{\bibinfo}[2]{#2}
\providecommand{\eprint}[2][]{\url{#2}}

\bibitem[{\citenamefont{Wen and Zee}(1993)}]{WenZee1993}
\bibinfo{author}{\bibfnamefont{X.~G.} \bibnamefont{Wen}} \bibnamefont{and}
  \bibinfo{author}{\bibfnamefont{A.}~\bibnamefont{Zee}},
  \bibinfo{journal}{Phys. Rev. B} \textbf{\bibinfo{volume}{47}},
  \bibinfo{pages}{2265} (\bibinfo{year}{1993}).

\bibitem[{\citenamefont{Moon et~al.}(1995)\citenamefont{Moon, Mori, Yang,
  Girvin, MacDonald, Zheng, Yoshioka, and Zhang}}]{Moon1995}
\bibinfo{author}{\bibfnamefont{K.}~\bibnamefont{Moon}},
  \bibinfo{author}{\bibfnamefont{H.}~\bibnamefont{Mori}},
  \bibinfo{author}{\bibfnamefont{K.}~\bibnamefont{Yang}},
  \bibinfo{author}{\bibfnamefont{S.~M.} \bibnamefont{Girvin}},
  \bibinfo{author}{\bibfnamefont{A.~H.} \bibnamefont{MacDonald}},
  \bibinfo{author}{\bibfnamefont{L.}~\bibnamefont{Zheng}},
  \bibinfo{author}{\bibfnamefont{D.}~\bibnamefont{Yoshioka}}, \bibnamefont{and}
  \bibinfo{author}{\bibfnamefont{S.-C.} \bibnamefont{Zhang}},
  \bibinfo{journal}{Phys. Rev. B} \textbf{\bibinfo{volume}{51}},
  \bibinfo{pages}{5138} (\bibinfo{year}{1995}).

\bibitem[{\citenamefont{Eisenstein and MacDonald}(2004)}]{Eisenstein2004}
\bibinfo{author}{\bibfnamefont{J.~P.} \bibnamefont{Eisenstein}}
  \bibnamefont{and} \bibinfo{author}{\bibfnamefont{A.~H.}
  \bibnamefont{MacDonald}}, \bibinfo{journal}{Nature}
  \textbf{\bibinfo{volume}{432}}, \bibinfo{pages}{691} (\bibinfo{year}{2004}).

\bibitem[{\citenamefont{Spielman et~al.}(2000)\citenamefont{Spielman,
  Eisenstein, Pfeiffer, and West}}]{Spielman2000}
\bibinfo{author}{\bibfnamefont{I.~B.} \bibnamefont{Spielman}},
  \bibinfo{author}{\bibfnamefont{J.~P.} \bibnamefont{Eisenstein}},
  \bibinfo{author}{\bibfnamefont{L.~N.} \bibnamefont{Pfeiffer}},
  \bibnamefont{and} \bibinfo{author}{\bibfnamefont{K.~W.} \bibnamefont{West}},
  \bibinfo{journal}{Phys. Rev. Lett.} \textbf{\bibinfo{volume}{84}},
  \bibinfo{pages}{5808} (\bibinfo{year}{2000}).

\bibitem[{\citenamefont{Spielman et~al.}(2001)\citenamefont{Spielman,
  Eisenstein, Pfeiffer, and West}}]{Spielman2001}
\bibinfo{author}{\bibfnamefont{I.~B.} \bibnamefont{Spielman}},
  \bibinfo{author}{\bibfnamefont{J.~P.} \bibnamefont{Eisenstein}},
  \bibinfo{author}{\bibfnamefont{L.~N.} \bibnamefont{Pfeiffer}},
  \bibnamefont{and} \bibinfo{author}{\bibfnamefont{K.~W.} \bibnamefont{West}},
  \bibinfo{journal}{Phys. Rev. Lett.} \textbf{\bibinfo{volume}{87}},
  \bibinfo{pages}{036803} (\bibinfo{year}{2001}).

\bibitem[{\citenamefont{Kellogg et~al.}(2002)\citenamefont{Kellogg, Spielman,
  Eisenstein, Pfeiffer, and West}}]{Kellogg2002}
\bibinfo{author}{\bibfnamefont{M.}~\bibnamefont{Kellogg}},
  \bibinfo{author}{\bibfnamefont{I.~B.} \bibnamefont{Spielman}},
  \bibinfo{author}{\bibfnamefont{J.~P.} \bibnamefont{Eisenstein}},
  \bibinfo{author}{\bibfnamefont{L.~N.} \bibnamefont{Pfeiffer}},
  \bibnamefont{and} \bibinfo{author}{\bibfnamefont{K.~W.} \bibnamefont{West}},
  \bibinfo{journal}{Phys. Rev. Lett.} \textbf{\bibinfo{volume}{88}},
  \bibinfo{pages}{126804} (\bibinfo{year}{2002}).

\bibitem[{\citenamefont{Kellogg et~al.}(2004)\citenamefont{Kellogg, Eisenstein,
  Pfeiffer, and West}}]{Kellogg2004}
\bibinfo{author}{\bibfnamefont{M.}~\bibnamefont{Kellogg}},
  \bibinfo{author}{\bibfnamefont{J.~P.} \bibnamefont{Eisenstein}},
  \bibinfo{author}{\bibfnamefont{L.~N.} \bibnamefont{Pfeiffer}},
  \bibnamefont{and} \bibinfo{author}{\bibfnamefont{K.~W.} \bibnamefont{West}},
  \bibinfo{journal}{Phys. Rev. Lett.} \textbf{\bibinfo{volume}{93}},
  \bibinfo{pages}{036801} (\bibinfo{year}{2004}).

\bibitem[{\citenamefont{Stern et~al.}(2001)\citenamefont{Stern, Girvin,
  MacDonald, and Ma}}]{Stern2001}
\bibinfo{author}{\bibfnamefont{A.}~\bibnamefont{Stern}},
  \bibinfo{author}{\bibfnamefont{S.~M.} \bibnamefont{Girvin}},
  \bibinfo{author}{\bibfnamefont{A.~H.} \bibnamefont{MacDonald}},
  \bibnamefont{and} \bibinfo{author}{\bibfnamefont{N.}~\bibnamefont{Ma}},
  \bibinfo{journal}{Phys. Rev. Lett.} \textbf{\bibinfo{volume}{86}},
  \bibinfo{pages}{1829} (\bibinfo{year}{2001}).

\bibitem[{\citenamefont{Joglekar and MacDonald}(2001)}]{MacDonald2001}
\bibinfo{author}{\bibfnamefont{Y.~N.} \bibnamefont{Joglekar}} \bibnamefont{and}
  \bibinfo{author}{\bibfnamefont{A.~H.} \bibnamefont{MacDonald}},
  \bibinfo{journal}{Phys. Rev. Lett.} \textbf{\bibinfo{volume}{87}},
  \bibinfo{pages}{196802} (\bibinfo{year}{2001}).

\bibitem[{\citenamefont{Fertig and Murthy}(2005)}]{Fertig2005}
\bibinfo{author}{\bibfnamefont{H.~A.} \bibnamefont{Fertig}} \bibnamefont{and}
  \bibinfo{author}{\bibfnamefont{G.}~\bibnamefont{Murthy}},
  \bibinfo{journal}{Phys. Rev. Lett.} \textbf{\bibinfo{volume}{95}},
  \bibinfo{pages}{156802} (\bibinfo{year}{2005}).

\bibitem[{\citenamefont{Halperin et~al.}(1993)\citenamefont{Halperin, Lee, and
  Read}}]{HalperinLeeRead}
\bibinfo{author}{\bibfnamefont{B.~I.} \bibnamefont{Halperin}},
  \bibinfo{author}{\bibfnamefont{P.~A.} \bibnamefont{Lee}}, \bibnamefont{and}
  \bibinfo{author}{\bibfnamefont{N.}~\bibnamefont{Read}},
  \bibinfo{journal}{Phys. Rev. B} \textbf{\bibinfo{volume}{47}},
  \bibinfo{pages}{7312} (\bibinfo{year}{1993}).

\bibitem[{\citenamefont{Simon and Halperin}(1993)}]{MRPA}
\bibinfo{author}{\bibfnamefont{S.~H.} \bibnamefont{Simon}} \bibnamefont{and}
  \bibinfo{author}{\bibfnamefont{B.~I.} \bibnamefont{Halperin}},
  \bibinfo{journal}{Phys. Rev. B} \textbf{\bibinfo{volume}{48}},
  \bibinfo{pages}{17368} (\bibinfo{year}{1993}).

\bibitem[{\citenamefont{Stern and Halperin}(1995)}]{SternHalperin1995}
\bibinfo{author}{\bibfnamefont{A.}~\bibnamefont{Stern}} \bibnamefont{and}
  \bibinfo{author}{\bibfnamefont{B.~I.} \bibnamefont{Halperin}},
  \bibinfo{journal}{Phys. Rev. B} \textbf{\bibinfo{volume}{52}},
  \bibinfo{pages}{5890} (\bibinfo{year}{1995}).

\bibitem[{\citenamefont{Simon et~al.}(1996)\citenamefont{Simon, Stern, and
  Halperin}}]{MMRPA}
\bibinfo{author}{\bibfnamefont{S.~H.} \bibnamefont{Simon}},
  \bibinfo{author}{\bibfnamefont{A.}~\bibnamefont{Stern}}, \bibnamefont{and}
  \bibinfo{author}{\bibfnamefont{B.~I.} \bibnamefont{Halperin}},
  \bibinfo{journal}{Phys. Rev. B} \textbf{\bibinfo{volume}{54}},
  \bibinfo{pages}{R11114} (\bibinfo{year}{1996}).

\bibitem[{\citenamefont{Simon}(1998)}]{Simon1998}
\bibinfo{author}{\bibfnamefont{S.~H.} \bibnamefont{Simon}}, in
  \emph{\bibinfo{booktitle}{Composite Fermions}}, edited by
  \bibinfo{editor}{\bibfnamefont{O.}~\bibnamefont{Heinonen}}
  (\bibinfo{publisher}{World Scientific}, \bibinfo{year}{1998}).

\bibitem[{\citenamefont{Shankar and Murthy}(1997)}]{Shankar1997}
\bibinfo{author}{\bibfnamefont{R.}~\bibnamefont{Shankar}} \bibnamefont{and}
  \bibinfo{author}{\bibfnamefont{G.}~\bibnamefont{Murthy}},
  \bibinfo{journal}{Phys. Rev. Lett.} \textbf{\bibinfo{volume}{79}},
  \bibinfo{pages}{4437} (\bibinfo{year}{1997}).

\bibitem[{\citenamefont{Read}(1998)}]{Read1998}
\bibinfo{author}{\bibfnamefont{N.}~\bibnamefont{Read}}, \bibinfo{journal}{Phys.
  Rev. B} \textbf{\bibinfo{volume}{58}}, \bibinfo{pages}{16262}
  (\bibinfo{year}{1998}).

\bibitem[{\citenamefont{Lee}(1998)}]{Lee1998}
\bibinfo{author}{\bibfnamefont{D.-H.} \bibnamefont{Lee}},
  \bibinfo{journal}{Phys. Rev. Lett.} \textbf{\bibinfo{volume}{80}},
  \bibinfo{pages}{4745} (\bibinfo{year}{1998}).

\bibitem[{\citenamefont{Stern et~al.}(1999)\citenamefont{Stern, Halperin, von
  Oppen, and Simon}}]{Stern1999}
\bibinfo{author}{\bibfnamefont{A.}~\bibnamefont{Stern}},
  \bibinfo{author}{\bibfnamefont{B.~I.} \bibnamefont{Halperin}},
  \bibinfo{author}{\bibfnamefont{F.}~\bibnamefont{von Oppen}},
  \bibnamefont{and} \bibinfo{author}{\bibfnamefont{S.~H.} \bibnamefont{Simon}},
  \bibinfo{journal}{Phys. Rev. B} \textbf{\bibinfo{volume}{59}},
  \bibinfo{pages}{12547} (\bibinfo{year}{1999}).

\bibitem[{\citenamefont{Shankar}(2001)}]{Shankar2001}
\bibinfo{author}{\bibfnamefont{R.}~\bibnamefont{Shankar}},
  \bibinfo{journal}{Phys. Rev. B} \textbf{\bibinfo{volume}{63}},
  \bibinfo{pages}{085322} (\bibinfo{year}{2001}).

\bibitem[{\citenamefont{Murthy and Shankar}(2003)}]{Shankar2003}
\bibinfo{author}{\bibfnamefont{G.}~\bibnamefont{Murthy}} \bibnamefont{and}
  \bibinfo{author}{\bibfnamefont{R.}~\bibnamefont{Shankar}},
  \bibinfo{journal}{Rev. Mod. Phys.} \textbf{\bibinfo{volume}{75}},
  \bibinfo{pages}{1101} (\bibinfo{year}{2003}).

\bibitem[{\citenamefont{Kellogg et~al.}(2003)\citenamefont{Kellogg, Eisenstein,
  Pfeiffer, and West}}]{Kellogg2003}
\bibinfo{author}{\bibfnamefont{M.}~\bibnamefont{Kellogg}},
  \bibinfo{author}{\bibfnamefont{J.~P.} \bibnamefont{Eisenstein}},
  \bibinfo{author}{\bibfnamefont{L.~N.} \bibnamefont{Pfeiffer}},
  \bibnamefont{and} \bibinfo{author}{\bibfnamefont{K.~W.} \bibnamefont{West}},
  \bibinfo{journal}{Phys. Rev. Lett.} \textbf{\bibinfo{volume}{90}},
  \bibinfo{pages}{246801} (\bibinfo{year}{2003}).

\bibitem[{\citenamefont{Tutuc et~al.}(2003)\citenamefont{Tutuc, Melinte,
  De~Poortere, Pillarisetty, and Shayegan}}]{Tutuc2003}
\bibinfo{author}{\bibfnamefont{E.}~\bibnamefont{Tutuc}},
  \bibinfo{author}{\bibfnamefont{S.}~\bibnamefont{Melinte}},
  \bibinfo{author}{\bibfnamefont{E.~P.} \bibnamefont{De~Poortere}},
  \bibinfo{author}{\bibfnamefont{R.}~\bibnamefont{Pillarisetty}},
  \bibnamefont{and} \bibinfo{author}{\bibfnamefont{M.}~\bibnamefont{Shayegan}},
  \bibinfo{journal}{Phys. Rev. Lett.} \textbf{\bibinfo{volume}{91}},
  \bibinfo{pages}{076802} (\bibinfo{year}{2003}).

\bibitem[{\citenamefont{Spielman et~al.}(2004)\citenamefont{Spielman, Kellogg,
  Eisenstein, Pfeiffer, and West}}]{Spielman2004}
\bibinfo{author}{\bibfnamefont{I.~B.} \bibnamefont{Spielman}},
  \bibinfo{author}{\bibfnamefont{M.}~\bibnamefont{Kellogg}},
  \bibinfo{author}{\bibfnamefont{J.~P.} \bibnamefont{Eisenstein}},
  \bibinfo{author}{\bibfnamefont{L.~N.} \bibnamefont{Pfeiffer}},
  \bibnamefont{and} \bibinfo{author}{\bibfnamefont{K.~W.} \bibnamefont{West}},
  \bibinfo{journal}{Phys. Rev. B} \textbf{\bibinfo{volume}{70}},
  \bibinfo{pages}{081303} (\bibinfo{year}{2004}).

\bibitem[{\citenamefont{Clarke et~al.}(2005)\citenamefont{Clarke, Micolich,
  Hamilton, Simmons, Hanna, Rodriguez, Pepper, and Ritchie}}]{Clarke2005}
\bibinfo{author}{\bibfnamefont{W.~R.} \bibnamefont{Clarke}},
  \bibinfo{author}{\bibfnamefont{A.~P.} \bibnamefont{Micolich}},
  \bibinfo{author}{\bibfnamefont{A.~R.} \bibnamefont{Hamilton}},
  \bibinfo{author}{\bibfnamefont{M.~Y.} \bibnamefont{Simmons}},
  \bibinfo{author}{\bibfnamefont{C.~B.} \bibnamefont{Hanna}},
  \bibinfo{author}{\bibfnamefont{J.~R.} \bibnamefont{Rodriguez}},
  \bibinfo{author}{\bibfnamefont{M.}~\bibnamefont{Pepper}}, \bibnamefont{and}
  \bibinfo{author}{\bibfnamefont{D.~A.} \bibnamefont{Ritchie}},
  \bibinfo{journal}{Phys. Rev. B} \textbf{\bibinfo{volume}{71}},
  \bibinfo{pages}{081304} (\bibinfo{year}{2005}).

\bibitem[{\citenamefont{Spielman et~al.}(2005)\citenamefont{Spielman, Tracy,
  Eisenstein, Pfeiffer, and West}}]{Spielman2005}
\bibinfo{author}{\bibfnamefont{I.~B.} \bibnamefont{Spielman}},
  \bibinfo{author}{\bibfnamefont{L.~A.} \bibnamefont{Tracy}},
  \bibinfo{author}{\bibfnamefont{J.~P.} \bibnamefont{Eisenstein}},
  \bibinfo{author}{\bibfnamefont{L.~N.} \bibnamefont{Pfeiffer}},
  \bibnamefont{and} \bibinfo{author}{\bibfnamefont{K.~W.} \bibnamefont{West}},
  \bibinfo{journal}{Phys. Rev. Lett.} \textbf{\bibinfo{volume}{94}},
  \bibinfo{pages}{076803} (\bibinfo{year}{2005}).

\bibitem[{\citenamefont{Kumada et~al.}(2005)\citenamefont{Kumada, Muraki,
  Hashimoto, and Hirayama}}]{Kumada2005}
\bibinfo{author}{\bibfnamefont{N.}~\bibnamefont{Kumada}},
  \bibinfo{author}{\bibfnamefont{K.}~\bibnamefont{Muraki}},
  \bibinfo{author}{\bibfnamefont{K.}~\bibnamefont{Hashimoto}},
  \bibnamefont{and} \bibinfo{author}{\bibfnamefont{Y.}~\bibnamefont{Hirayama}},
  \bibinfo{journal}{Phys. Rev. Lett.} \textbf{\bibinfo{volume}{94}},
  \bibinfo{pages}{096802} (\bibinfo{year}{2005}).

\bibitem[{\citenamefont{Luin et~al.}(2005)\citenamefont{Luin, Pellegrini,
  Pinczuk, Dennis, Pfeiffer, and West}}]{Luin2005}
\bibinfo{author}{\bibfnamefont{S.}~\bibnamefont{Luin}},
  \bibinfo{author}{\bibfnamefont{V.}~\bibnamefont{Pellegrini}},
  \bibinfo{author}{\bibfnamefont{A.}~\bibnamefont{Pinczuk}},
  \bibinfo{author}{\bibfnamefont{B.~S.} \bibnamefont{Dennis}},
  \bibinfo{author}{\bibfnamefont{L.~N.} \bibnamefont{Pfeiffer}},
  \bibnamefont{and} \bibinfo{author}{\bibfnamefont{K.~W.} \bibnamefont{West}},
  \bibinfo{journal}{Phys. Rev. Lett.} \textbf{\bibinfo{volume}{94}},
  \bibinfo{pages}{146804} (\bibinfo{year}{2005}).

\bibitem[{\citenamefont{Tracy et~al.}(2006)\citenamefont{Tracy, Eisenstein,
  Pfeiffer, and West}}]{Tracy2006}
\bibinfo{author}{\bibfnamefont{L.~A.} \bibnamefont{Tracy}},
  \bibinfo{author}{\bibfnamefont{J.~P.} \bibnamefont{Eisenstein}},
  \bibinfo{author}{\bibfnamefont{L.~N.} \bibnamefont{Pfeiffer}},
  \bibnamefont{and} \bibinfo{author}{\bibfnamefont{K.~W.} \bibnamefont{West}},
  \bibinfo{journal}{Phys. Rev. B} \textbf{\bibinfo{volume}{73}},
  \bibinfo{pages}{121306} (\bibinfo{year}{2006}).

\bibitem[{\citenamefont{Giudici et~al.}(2008)\citenamefont{Giudici, Muraki,
  Kumada, Hirayama, and Fujisawa}}]{Fujisawa2008}
\bibinfo{author}{\bibfnamefont{P.}~\bibnamefont{Giudici}},
  \bibinfo{author}{\bibfnamefont{K.}~\bibnamefont{Muraki}},
  \bibinfo{author}{\bibfnamefont{N.}~\bibnamefont{Kumada}},
  \bibinfo{author}{\bibfnamefont{Y.}~\bibnamefont{Hirayama}}, \bibnamefont{and}
  \bibinfo{author}{\bibfnamefont{T.}~\bibnamefont{Fujisawa}},
  \bibinfo{journal}{Phys. Rev. Lett.} \textbf{\bibinfo{volume}{100}},
  \bibinfo{pages}{106803} (\bibinfo{year}{2008}).

\bibitem[{\citenamefont{Champagne
  et~al.}(2008{\natexlab{a}})\citenamefont{Champagne, Eisenstein, Pfeiffer, and
  West}}]{Champagne2008}
\bibinfo{author}{\bibfnamefont{A.~R.} \bibnamefont{Champagne}},
  \bibinfo{author}{\bibfnamefont{J.~P.} \bibnamefont{Eisenstein}},
  \bibinfo{author}{\bibfnamefont{L.~N.} \bibnamefont{Pfeiffer}},
  \bibnamefont{and} \bibinfo{author}{\bibfnamefont{K.~W.} \bibnamefont{West}},
  \bibinfo{journal}{Phys. Rev. Lett.} \textbf{\bibinfo{volume}{100}}
  (\bibinfo{year}{2008}{\natexlab{a}}).

\bibitem[{\citenamefont{Champagne
  et~al.}(2008{\natexlab{b}})\citenamefont{Champagne, Finck, Eisenstein,
  Pfeiffer, and West}}]{Champagne2008b}
\bibinfo{author}{\bibfnamefont{A.~R.} \bibnamefont{Champagne}},
  \bibinfo{author}{\bibfnamefont{A.~D.~K.} \bibnamefont{Finck}},
  \bibinfo{author}{\bibfnamefont{J.~P.} \bibnamefont{Eisenstein}},
  \bibinfo{author}{\bibfnamefont{L.~N.} \bibnamefont{Pfeiffer}},
  \bibnamefont{and} \bibinfo{author}{\bibfnamefont{K.~W.} \bibnamefont{West}},
  \bibinfo{journal}{Phys. Rev. B} \textbf{\bibinfo{volume}{78}},
  \bibinfo{pages}{205310} (\bibinfo{year}{2008}{\natexlab{b}}).

\bibitem[{\citenamefont{Karmakar et~al.}(2009)\citenamefont{Karmakar,
  Pellegrini, Pinczuk, Pfeiffer, and West}}]{Karmakar2009}
\bibinfo{author}{\bibfnamefont{B.}~\bibnamefont{Karmakar}},
  \bibinfo{author}{\bibfnamefont{V.}~\bibnamefont{Pellegrini}},
  \bibinfo{author}{\bibfnamefont{A.}~\bibnamefont{Pinczuk}},
  \bibinfo{author}{\bibfnamefont{L.~N.} \bibnamefont{Pfeiffer}},
  \bibnamefont{and} \bibinfo{author}{\bibfnamefont{K.~W.} \bibnamefont{West}},
  \bibinfo{journal}{Phys. Rev. Lett.} \textbf{\bibinfo{volume}{102}},
  \bibinfo{pages}{036802} (\bibinfo{year}{2009}).

\bibitem[{\citenamefont{Finck et~al.}(2010)\citenamefont{Finck, Eisenstein,
  Pfeiffer, and West}}]{Finck2010}
\bibinfo{author}{\bibfnamefont{A.~D.~K.} \bibnamefont{Finck}},
  \bibinfo{author}{\bibfnamefont{J.~P.} \bibnamefont{Eisenstein}},
  \bibinfo{author}{\bibfnamefont{L.~N.} \bibnamefont{Pfeiffer}},
  \bibnamefont{and} \bibinfo{author}{\bibfnamefont{K.~W.} \bibnamefont{West}},
  \bibinfo{journal}{Phys. Rev. Lett.} \textbf{\bibinfo{volume}{104}},
  \bibinfo{pages}{016801} (\bibinfo{year}{2010}).

\bibitem[{\citenamefont{Giudici et~al.}(2010)\citenamefont{Giudici, Muraki,
  Kumada, and Fujisawa}}]{Muraki2010}
\bibinfo{author}{\bibfnamefont{P.}~\bibnamefont{Giudici}},
  \bibinfo{author}{\bibfnamefont{K.}~\bibnamefont{Muraki}},
  \bibinfo{author}{\bibfnamefont{N.}~\bibnamefont{Kumada}}, \bibnamefont{and}
  \bibinfo{author}{\bibfnamefont{T.}~\bibnamefont{Fujisawa}},
  \bibinfo{journal}{Phys. Rev. Lett.} \textbf{\bibinfo{volume}{104}},
  \bibinfo{pages}{056802} (\bibinfo{year}{2010}).

\bibitem[{\citenamefont{C\^ot\'e et~al.}(1992)\citenamefont{C\^ot\'e, Brey, and
  MacDonald}}]{Cote1992}
\bibinfo{author}{\bibfnamefont{R.}~\bibnamefont{C\^ot\'e}},
  \bibinfo{author}{\bibfnamefont{L.}~\bibnamefont{Brey}}, \bibnamefont{and}
  \bibinfo{author}{\bibfnamefont{A.~H.} \bibnamefont{MacDonald}},
  \bibinfo{journal}{Phys. Rev. B} \textbf{\bibinfo{volume}{46}},
  \bibinfo{pages}{10239} (\bibinfo{year}{1992}).

\bibitem[{\citenamefont{Zheng and Fertig}(1995)}]{ZhengFertig1995}
\bibinfo{author}{\bibfnamefont{L.}~\bibnamefont{Zheng}} \bibnamefont{and}
  \bibinfo{author}{\bibfnamefont{H.~A.} \bibnamefont{Fertig}},
  \bibinfo{journal}{Phys. Rev. B} \textbf{\bibinfo{volume}{52}},
  \bibinfo{pages}{12282} (\bibinfo{year}{1995}).

\bibitem[{\citenamefont{Narasimhan and Ho}(1995)}]{Ho1995}
\bibinfo{author}{\bibfnamefont{S.}~\bibnamefont{Narasimhan}} \bibnamefont{and}
  \bibinfo{author}{\bibfnamefont{T.-L.} \bibnamefont{Ho}},
  \bibinfo{journal}{Phys. Rev. B} \textbf{\bibinfo{volume}{52}},
  \bibinfo{pages}{12291} (\bibinfo{year}{1995}).

\bibitem[{\citenamefont{Bonesteel et~al.}(1996)\citenamefont{Bonesteel,
  McDonald, and Nayak}}]{Bonesteel1996}
\bibinfo{author}{\bibfnamefont{N.~E.} \bibnamefont{Bonesteel}},
  \bibinfo{author}{\bibfnamefont{I.~A.} \bibnamefont{McDonald}},
  \bibnamefont{and} \bibinfo{author}{\bibfnamefont{C.}~\bibnamefont{Nayak}},
  \bibinfo{journal}{Phys. Rev. Lett.} \textbf{\bibinfo{volume}{77}},
  \bibinfo{pages}{3009} (\bibinfo{year}{1996}).

\bibitem[{\citenamefont{Kim et~al.}(2001)\citenamefont{Kim, Nayak, Demler,
  Read, and Das~Sarma}}]{Kim2001}
\bibinfo{author}{\bibfnamefont{Y.~B.} \bibnamefont{Kim}},
  \bibinfo{author}{\bibfnamefont{C.}~\bibnamefont{Nayak}},
  \bibinfo{author}{\bibfnamefont{E.}~\bibnamefont{Demler}},
  \bibinfo{author}{\bibfnamefont{N.}~\bibnamefont{Read}}, \bibnamefont{and}
  \bibinfo{author}{\bibfnamefont{S.}~\bibnamefont{Das~Sarma}},
  \bibinfo{journal}{Phys. Rev. B} \textbf{\bibinfo{volume}{63}},
  \bibinfo{pages}{205315} (\bibinfo{year}{2001}).

\bibitem[{\citenamefont{Demler et~al.}(2001)\citenamefont{Demler, Nayak, and
  Das~Sarma}}]{Demler2001}
\bibinfo{author}{\bibfnamefont{E.}~\bibnamefont{Demler}},
  \bibinfo{author}{\bibfnamefont{C.}~\bibnamefont{Nayak}}, \bibnamefont{and}
  \bibinfo{author}{\bibfnamefont{S.}~\bibnamefont{Das~Sarma}},
  \bibinfo{journal}{Phys. Rev. Lett.} \textbf{\bibinfo{volume}{86}},
  \bibinfo{pages}{1853} (\bibinfo{year}{2001}).

\bibitem[{\citenamefont{Schliemann et~al.}(2001)\citenamefont{Schliemann,
  Girvin, and MacDonald}}]{Schliemann2001}
\bibinfo{author}{\bibfnamefont{J.}~\bibnamefont{Schliemann}},
  \bibinfo{author}{\bibfnamefont{S.~M.} \bibnamefont{Girvin}},
  \bibnamefont{and} \bibinfo{author}{\bibfnamefont{A.~H.}
  \bibnamefont{MacDonald}}, \bibinfo{journal}{Phys. Rev. Lett.}
  \textbf{\bibinfo{volume}{86}}, \bibinfo{pages}{1849} (\bibinfo{year}{2001}).

\bibitem[{\citenamefont{Stern and Halperin}(2002)}]{SternHalperin2001}
\bibinfo{author}{\bibfnamefont{A.}~\bibnamefont{Stern}} \bibnamefont{and}
  \bibinfo{author}{\bibfnamefont{B.~I.} \bibnamefont{Halperin}},
  \bibinfo{journal}{Phys. Rev. Lett.} \textbf{\bibinfo{volume}{88}},
  \bibinfo{pages}{106801} (\bibinfo{year}{2002}).

\bibitem[{\citenamefont{Simon et~al.}(2003)\citenamefont{Simon, Rezayi, and
  Milovanovic}}]{Simon2003}
\bibinfo{author}{\bibfnamefont{S.~H.} \bibnamefont{Simon}},
  \bibinfo{author}{\bibfnamefont{E.~H.} \bibnamefont{Rezayi}},
  \bibnamefont{and} \bibinfo{author}{\bibfnamefont{M.~V.}
  \bibnamefont{Milovanovic}}, \bibinfo{journal}{Phys. Rev. Lett.}
  \textbf{\bibinfo{volume}{91}}, \bibinfo{pages}{046803}
  (\bibinfo{year}{2003}).

\bibitem[{\citenamefont{Shibata and Yoshioka}(2006)}]{Shibata2006}
\bibinfo{author}{\bibfnamefont{N.}~\bibnamefont{Shibata}} \bibnamefont{and}
  \bibinfo{author}{\bibfnamefont{D.}~\bibnamefont{Yoshioka}},
  \bibinfo{journal}{J. Phys. Soc. Jpn.} \textbf{\bibinfo{volume}{75}},
  \bibinfo{pages}{043712} (\bibinfo{year}{2006}).

\bibitem[{\citenamefont{Ye and Jiang}(2007)}]{Ye2007}
\bibinfo{author}{\bibfnamefont{J.}~\bibnamefont{Ye}} \bibnamefont{and}
  \bibinfo{author}{\bibfnamefont{L.}~\bibnamefont{Jiang}},
  \bibinfo{journal}{Phys. Rev. Lett.} \textbf{\bibinfo{volume}{98}},
  \bibinfo{pages}{236802} (\bibinfo{year}{2007}).

\bibitem[{\citenamefont{M\"{o}ller et~al.}(2008)\citenamefont{M\"{o}ller,
  Simon, and Rezayi}}]{Moller2008a}
\bibinfo{author}{\bibfnamefont{G.}~\bibnamefont{M\"{o}ller}},
  \bibinfo{author}{\bibfnamefont{S.~H.} \bibnamefont{Simon}}, \bibnamefont{and}
  \bibinfo{author}{\bibfnamefont{E.~H.} \bibnamefont{Rezayi}},
  \bibinfo{journal}{Phys. Rev. Lett.} \textbf{\bibinfo{volume}{101}},
  \bibinfo{pages}{176803} (\bibinfo{year}{2008}).

\bibitem[{\citenamefont{Moller and Simon}(2008)}]{Moller2008}
\bibinfo{author}{\bibfnamefont{G.}~\bibnamefont{M\"{o}ller}} \bibnamefont{and}
  \bibinfo{author}{\bibfnamefont{S.~H.} \bibnamefont{Simon}},
  \bibinfo{journal}{Phys.Rev. B} \textbf{\bibinfo{volume}{77}},
  \bibinfo{pages}{075319} (\bibinfo{year}{2008}).

\bibitem[{\citenamefont{Milovanovi\ifmmode~\acute{c}\else \'{c}\fi{} and
  Papi\ifmmode~\acute{c}\else \'{c}\fi{}}(2009)}]{Papic2009}
\bibinfo{author}{\bibfnamefont{M.~V.}
  \bibnamefont{Milovanovi\ifmmode~\acute{c}\else \'{c}\fi{}}} \bibnamefont{and}
  \bibinfo{author}{\bibfnamefont{Z.}~\bibnamefont{Papi\ifmmode~\acute{c}\else
  \'{c}\fi{}}}, \bibinfo{journal}{Phys. Rev. B} \textbf{\bibinfo{volume}{79}},
  \bibinfo{pages}{115319} (\bibinfo{year}{2009}).

\bibitem[{\citenamefont{Tracy et~al.}(2007)\citenamefont{Tracy, Eisenstein,
  Pfeiffer, and West}}]{Tracy2007}
\bibinfo{author}{\bibfnamefont{L.~A.} \bibnamefont{Tracy}},
  \bibinfo{author}{\bibfnamefont{J.~P.} \bibnamefont{Eisenstein}},
  \bibinfo{author}{\bibfnamefont{L.~N.} \bibnamefont{Pfeiffer}},
  \bibnamefont{and} \bibinfo{author}{\bibfnamefont{K.~W.} \bibnamefont{West}},
  \bibinfo{journal}{Phys. Rev. Lett.} \textbf{\bibinfo{volume}{98}},
  \bibinfo{pages}{086801} (\bibinfo{year}{2007}).

\bibitem[{\citenamefont{Li et~al.}(2009)\citenamefont{Li, Umansky, von
  Klitzing, and Smet}}]{Li2009}
\bibinfo{author}{\bibfnamefont{Y.~Q.} \bibnamefont{Li}},
  \bibinfo{author}{\bibfnamefont{V.}~\bibnamefont{Umansky}},
  \bibinfo{author}{\bibfnamefont{K.}~\bibnamefont{von Klitzing}},
  \bibnamefont{and} \bibinfo{author}{\bibfnamefont{J.~H.} \bibnamefont{Smet}},
  \bibinfo{journal}{Phys. Rev. Lett.} \textbf{\bibinfo{volume}{102}},
  \bibinfo{pages}{046803} (\bibinfo{year}{2009}).

\bibitem[{\citenamefont{Park and Jain}(1998{\natexlab{a}})}]{ParkJain1998}
\bibinfo{author}{\bibfnamefont{K.}~\bibnamefont{Park}} \bibnamefont{and}
  \bibinfo{author}{\bibfnamefont{J.~K.} \bibnamefont{Jain}},
  \bibinfo{journal}{Phys. Rev. Lett.} \textbf{\bibinfo{volume}{80}},
  \bibinfo{pages}{4237} (\bibinfo{year}{1998}{\natexlab{a}}).

\bibitem[{\citenamefont{Melinte et~al.}(2000)\citenamefont{Melinte, Freytag,
  Horvati\ifmmode~\acute{c}\else \'{c}\fi{}, Berthier, L\'evy, Bayot, and
  Shayegan}}]{Melinte2000}
\bibinfo{author}{\bibfnamefont{S.}~\bibnamefont{Melinte}},
  \bibinfo{author}{\bibfnamefont{N.}~\bibnamefont{Freytag}},
  \bibinfo{author}{\bibfnamefont{M.}~\bibnamefont{Horvati\ifmmode~\acute{c}\el%
se \'{c}\fi{}}}, \bibinfo{author}{\bibfnamefont{C.}~\bibnamefont{Berthier}},
  \bibinfo{author}{\bibfnamefont{L.~P.} \bibnamefont{L\'evy}},
  \bibinfo{author}{\bibfnamefont{V.}~\bibnamefont{Bayot}}, \bibnamefont{and}
  \bibinfo{author}{\bibfnamefont{M.}~\bibnamefont{Shayegan}},
  \bibinfo{journal}{Phys. Rev. Lett.} \textbf{\bibinfo{volume}{84}},
  \bibinfo{pages}{354} (\bibinfo{year}{2000}).

\bibitem[{\citenamefont{Freytag et~al.}(2002)\citenamefont{Freytag,
  Horvati\ifmmode~\acute{c}\else \'{c}\fi{}, Berthier, Shayegan, and
  L\'evy}}]{Freytag2002}
\bibinfo{author}{\bibfnamefont{N.}~\bibnamefont{Freytag}},
  \bibinfo{author}{\bibfnamefont{M.}~\bibnamefont{Horvati\ifmmode~\acute{c}\el%
se \'{c}\fi{}}}, \bibinfo{author}{\bibfnamefont{C.}~\bibnamefont{Berthier}},
  \bibinfo{author}{\bibfnamefont{M.}~\bibnamefont{Shayegan}}, \bibnamefont{and}
  \bibinfo{author}{\bibfnamefont{L.~P.} \bibnamefont{L\'evy}},
  \bibinfo{journal}{Phys. Rev. Lett.} \textbf{\bibinfo{volume}{89}},
  \bibinfo{pages}{246804} (\bibinfo{year}{2002}).

\bibitem[{\citenamefont{Kukushkin et~al.}(1999)\citenamefont{Kukushkin,
  v.~Klitzing, and Eberl}}]{Kukushkin1999}
\bibinfo{author}{\bibfnamefont{I.~V.} \bibnamefont{Kukushkin}},
  \bibinfo{author}{\bibfnamefont{K.}~\bibnamefont{v.~Klitzing}},
  \bibnamefont{and} \bibinfo{author}{\bibfnamefont{K.}~\bibnamefont{Eberl}},
  \bibinfo{journal}{Phys. Rev. Lett.} \textbf{\bibinfo{volume}{82}},
  \bibinfo{pages}{3665} (\bibinfo{year}{1999}).

\bibitem[{\citenamefont{Kim and Lee}(1996)}]{Kim1996}
\bibinfo{author}{\bibfnamefont{Y.~B.} \bibnamefont{Kim}} \bibnamefont{and}
  \bibinfo{author}{\bibfnamefont{P.~A.} \bibnamefont{Lee}},
  \bibinfo{journal}{Phys. Rev. B} \textbf{\bibinfo{volume}{54}},
  \bibinfo{pages}{2715} (\bibinfo{year}{1996}).

\bibitem[{\citenamefont{Price}(1984)}]{Price1984}
\bibinfo{author}{\bibfnamefont{P.~J.} \bibnamefont{Price}},
  \bibinfo{journal}{Phys. Rev. B} \textbf{\bibinfo{volume}{30}},
  \bibinfo{pages}{2234} (\bibinfo{year}{1984}).

\bibitem[{\citenamefont{Gold}(1987)}]{Gold1987}
\bibinfo{author}{\bibfnamefont{A.}~\bibnamefont{Gold}}, \bibinfo{journal}{Phys.
  Rev. B} \textbf{\bibinfo{volume}{35}}, \bibinfo{pages}{723}
  (\bibinfo{year}{1987}).

\bibitem{\citenamefont{note1}}
\bibinfo{howpublished}{Since the gapless Goldstone mode contribution to $\partial E/\partial T$ is smaller than 2\% of the composite fermion contribution, even if gap of the topological excitations is comparable to the temperature and therefore the topological excitation contribution to $\partial E/\partial T$ is comparable to that of the gapless Goldstone mode, the composite fermion contribution would still dominate $\partial E/\partial T$ and our analysis is not affected.}

\bibitem[{\citenamefont{Wen}(2007)}]{Wen}
\bibinfo{author}{\bibfnamefont{X.~G.} \bibnamefont{Wen}},
  \emph{\bibinfo{title}{Quantum Field Theory of Many-body Systems: From the
  Origin of Sound to an Origin of Light and Electrons}}
  (\bibinfo{publisher}{Oxford University Press}, \bibinfo{year}{2007}).

\bibitem[{\citenamefont{Du et~al.}(1993)\citenamefont{Du, Stormer, Tsui,
  Pfeiffer, and West}}]{Du1993}
\bibinfo{author}{\bibfnamefont{R.~R.} \bibnamefont{Du}},
  \bibinfo{author}{\bibfnamefont{H.~L.} \bibnamefont{Stormer}},
  \bibinfo{author}{\bibfnamefont{D.~C.} \bibnamefont{Tsui}},
  \bibinfo{author}{\bibfnamefont{L.~N.} \bibnamefont{Pfeiffer}},
  \bibnamefont{and} \bibinfo{author}{\bibfnamefont{K.~W.} \bibnamefont{West}},
  \bibinfo{journal}{Phys. Rev. Lett.} \textbf{\bibinfo{volume}{70}},
  \bibinfo{pages}{2944} (\bibinfo{year}{1993}).

\bibitem[{\citenamefont{Du et~al.}(1994)\citenamefont{Du, Stormer, Tsui, Yeh,
  Pfeiffer, and West}}]{Du1994}
\bibinfo{author}{\bibfnamefont{R.~R.} \bibnamefont{Du}},
  \bibinfo{author}{\bibfnamefont{H.~L.} \bibnamefont{Stormer}},
  \bibinfo{author}{\bibfnamefont{D.~C.} \bibnamefont{Tsui}},
  \bibinfo{author}{\bibfnamefont{A.~S.} \bibnamefont{Yeh}},
  \bibinfo{author}{\bibfnamefont{L.~N.} \bibnamefont{Pfeiffer}},
  \bibnamefont{and} \bibinfo{author}{\bibfnamefont{K.~W.} \bibnamefont{West}},
  \bibinfo{journal}{Phys. Rev. Lett.} \textbf{\bibinfo{volume}{73}},
  \bibinfo{pages}{3274} (\bibinfo{year}{1994}).

\bibitem[{\citenamefont{Morf and d'Ambrumenil}(1995)}]{Morf1995}
\bibinfo{author}{\bibfnamefont{R.}~\bibnamefont{Morf}} \bibnamefont{and}
  \bibinfo{author}{\bibfnamefont{N.}~\bibnamefont{d'Ambrumenil}},
  \bibinfo{journal}{Phys. Rev. Lett.} \textbf{\bibinfo{volume}{74}},
  \bibinfo{pages}{5116} (\bibinfo{year}{1995}).

\bibitem[{\citenamefont{Park and Jain}(1998{\natexlab{b}})}]{ParkJain1999}
\bibinfo{author}{\bibfnamefont{K.}~\bibnamefont{Park}} \bibnamefont{and}
  \bibinfo{author}{\bibfnamefont{J.~K.} \bibnamefont{Jain}},
  \bibinfo{journal}{Phys. Rev. Lett.} \textbf{\bibinfo{volume}{81}},
  \bibinfo{pages}{4200} (\bibinfo{year}{1998}{\natexlab{b}}).

\bibitem[{\citenamefont{Yoshioka}(1986)}]{Yoshioka1986}
\bibinfo{author}{\bibfnamefont{D.}~\bibnamefont{Yoshioka}},
  \bibinfo{journal}{J. Phys. Soc. Jpn.} \textbf{\bibinfo{volume}{55}},
  \bibinfo{pages}{885} (\bibinfo{year}{1986}).

\bibitem[{\citenamefont{Morf}(1999)}]{Morf1999}
\bibinfo{author}{\bibfnamefont{R.~H.} \bibnamefont{Morf}},
  \bibinfo{journal}{Phys. Rev. Lett.} \textbf{\bibinfo{volume}{83}},
  \bibinfo{pages}{1485} (\bibinfo{year}{1999}).

\bibitem[{\citenamefont{Park et~al.}(1999)\citenamefont{Park, Meskini, and
  Jain}}]{ParkMeskiniJain1999}
\bibinfo{author}{\bibfnamefont{K.}~\bibnamefont{Park}},
  \bibinfo{author}{\bibfnamefont{N.}~\bibnamefont{Meskini}}, \bibnamefont{and}
  \bibinfo{author}{\bibfnamefont{J.~K.} \bibnamefont{Jain}},
  \bibinfo{journal}{J. Phys.: Condensed Matter} \textbf{\bibinfo{volume}{11}},
  \bibinfo{pages}{7283} (\bibinfo{year}{1999}).

\bibitem[{\citenamefont{Park et~al.}(1998)\citenamefont{Park, Melik-Alaverdian,
  Bonesteel, and Jain}}]{Park1998}
\bibinfo{author}{\bibfnamefont{K.}~\bibnamefont{Park}},
  \bibinfo{author}{\bibfnamefont{V.}~\bibnamefont{Melik-Alaverdian}},
  \bibinfo{author}{\bibfnamefont{N.~E.} \bibnamefont{Bonesteel}},
  \bibnamefont{and} \bibinfo{author}{\bibfnamefont{J.~K.} \bibnamefont{Jain}},
  \bibinfo{journal}{Phys. Rev. B} \textbf{\bibinfo{volume}{58}},
  \bibinfo{pages}{R10167} (\bibinfo{year}{1998}).

\bibitem[{\citenamefont{Eisenstein et~al.}(1994)\citenamefont{Eisenstein,
  Pfeiffer, and West}}]{Eisenstein1994}
\bibinfo{author}{\bibfnamefont{J.~P.} \bibnamefont{Eisenstein}},
  \bibinfo{author}{\bibfnamefont{L.~N.} \bibnamefont{Pfeiffer}},
  \bibnamefont{and} \bibinfo{author}{\bibfnamefont{K.~W.} \bibnamefont{West}},
  \bibinfo{journal}{Phys. Rev. B} \textbf{\bibinfo{volume}{50}},
  \bibinfo{pages}{1760} (\bibinfo{year}{1994}).

\end{thebibliography}

\end{document}